\documentclass[10pt,letter,twocolumn]{article}

\usepackage[letterpaper,margin=1.5cm]{geometry}

\usepackage[T1]{fontenc}
\usepackage[utf8]{inputenc}
\usepackage{amsmath,amssymb,graphicx,booktabs,mathptmx}
\usepackage{mathtools,etoolbox}
\allowdisplaybreaks[1]
\usepackage{placeins,caption}
\usepackage{microtype,needspace,balance}
\usepackage{xcolor}
\definecolor{RevisionRed}{RGB}{190,0,0}
\newif\ifrevisionred
\revisionredfalse
\DeclareRobustCommand{\rev}[1]{{\ifrevisionred\color{RevisionRed}%
\hypersetup{linkcolor=RevisionRed,citecolor=RevisionRed,urlcolor=RevisionRed}\fi#1}}

\ifrevisionred\fi
\usepackage[colorlinks=true,linkcolor=blue,citecolor=blue,urlcolor=blue]{hyperref}
\let\originaleqref\eqref
\renewcommand{\eqref}[1]{\rev{\originaleqref{#1}}}
\hypersetup{pdftitle={Birefringent imaging and interferometry with periodic orbits and accretion in Einstein-nonlinear-electrodynamics},pdfauthor={Shi-Hai Dong, Farokhnaz Hosseinifar, Orlando Luongo, Hassan Hassanabadi}}
\makeatletter
\patchcmd{\@BAlancecol}{\setbox\@leftcolumn}{\advance\@tempdima by 3pt\setbox\@leftcolumn}{}{\errmessage{Reference column balancing patch failed}}
\renewcommand{\tagform@}[1]{\maketag@@@{\ifrevisionred\color{RevisionRed}\fi(\ignorespaces#1\unskip\@@italiccorr)}}
\renewcommand\section{\@startsection{section}{1}{\z@}{-2.0ex plus -.5ex minus -.2ex}{1.0ex plus .2ex}{\normalfont\large}}
\renewcommand\subsection{\@startsection{subsection}{2}{\z@}{-1.6ex plus -.4ex minus -.2ex}{.7ex plus .2ex}{\normalfont\normalsize}}

\makeatother
\AtBeginDocument{%
\setlength{\abovedisplayskip}{8pt plus 2pt minus 2pt}%
\setlength{\belowdisplayskip}{8pt plus 2pt minus 2pt}%
\setlength{\abovedisplayshortskip}{0pt plus 2pt}%
\setlength{\belowdisplayshortskip}{5pt plus 2pt minus 2pt}%
\setlength{\jot}{2pt}}
\begin{document}
\setcounter{footnote}{0}

\twocolumn[
\begin{center}
\begin{minipage}{0.98\textwidth}
\centering
{\Large\bfseries
Birefringent Imaging and Interferometry with Periodic Orbits and Accretion in Einstein-nonlinear-electrodynamics\par}
\vspace{10pt}
{\normalsize
\mbox{Shi-Hai Dong\textsuperscript{1,2}\footnotemark[1],\enspace
Farokhnaz Hosseinifar\textsuperscript{3}\footnotemark[2],\enspace
Orlando Luongo\textsuperscript{4,5,6}\footnotemark[3],\enspace
Hassan Hassanabadi\textsuperscript{7,8}\footnotemark[4]}\par}
\vspace{9pt}
{\small\renewcommand{\arraystretch}{1.08}
\begin{tabular}{@{}r@{\enspace}l@{}}
\textsuperscript{1} & Research Center for Quantum Physics, Huzhou Normal University, Huzhou 313000, P. R. China\\
\textsuperscript{2} & CIC, Instituto Polit\'ecnico Nacional, UPALM, CDMX 07700, Mexico\\
\textsuperscript{3} & Center for Theoretical Physics, Khazar University, 41 Mehseti Street, Baku AZ-1096, Azerbaijan\\
\textsuperscript{4} & Istituto Nazionale di Fisica Nucleare, Sezione di Perugia, Perugia 06123, Italy\\
\textsuperscript{5} & SUNY Polytechnic Institute, 13502 Utica, New York, USA\\
\textsuperscript{6} & Al-Farabi Kazakh National University, Al-Farabi Ave. 71, 050040 Almaty, Kazakhstan\\
\textsuperscript{7} & Physics Department, California State University, Fresno, CA 93740, USA\\
\textsuperscript{8} & Khazar University, Department of Physics and Electronics, 41 Mahsati St., AZ-1096, Baku, Azerbaijan
\end{tabular}\par}
\vspace{4pt}
\end{minipage}
\end{center}
]

\footnotetext[1]{Email: \texttt{dongsh2@yahoo.com}}
\footnotetext[2]{Email: \texttt{f.hoseinifar94@gmail.com}}
\footnotetext[3]{Email: \texttt{orlando.luongo@unicam.it}}
\footnotetext[4]{Email: \texttt{hassanhassanabadi@mail.fresnostate.edu}}
\begin{abstract}

We investigate birefringent imaging, interferometric visibility, neutral periodic geodesics, and thin-disk radiation around a magnetically charged black hole in Einstein-nonlinear-electrodynamics, using the Euler-Heisenberg-type model of de Paula et al.~\cite{dePaula}. The nonlinear coupling is treated as a free model parameter. Two effective photon metrics give rise to distinct unstable light rings, capture thresholds, and images. Transforming their image-order intensities into signed Fourier visibilities reveals separate critical-curve oscillations and a narrow-ring modulation scale in the unresolved sum. This two-mode correspondence does not require photon motion between two potential barriers. Rational orbital frequencies are then used to classify neutral bound trajectories and provide conservative gravitational-wave snapshots relevant to extreme-mass-ratio inspirals. A stationary thin-disk calculation supplies flux, temperature, emitted efficiency, and polarization-resolved thermal spectra. Across the sampled parameters, magnetic charge dominates the changes, while matched Maxwell comparisons resolve smaller nonlinear corrections. The static, optically thin emitters and the orbiting, optically thick thermal disk are distinct source configurations. Polarization comparisons use the same material source within each configuration; their brightness and spectral results are conditional on that configuration.

\end{abstract}

\section{Introduction}

General relativity describes gravitation through spacetime curvature and predicts black holes as regions bounded by an event horizon. The Schwarzschild and Kerr solutions establish the nonrotating and rotating vacuum geometries, while gravitational-wave observations of binary black hole mergers provide direct evidence for their dynamical strong-field behavior~\cite{Einstein,Schwarzschild,Kerr,Abbott}. Black holes consequently offer a common setting in which orbital motion, photon trajectories, and accretion radiation can test the relation between geometry and observable signals. In the presence of nonlinear electromagnetic fields, this relation is especially subtle: the geometry governing neutral matter and the characteristic geometries governing photons need not coincide.

Nonlinear electrodynamics (NED) extends the Maxwell description by allowing the electromagnetic response to depend on field invariants. Its development includes the Born-Infeld theory, the quantum-electrodynamic effective action of Heisenberg and Euler, and Schwinger's treatment of vacuum polarization~\cite{BornInfeld,HeisenbergEuler,Schwinger}. Coupling nonlinear fields to gravity also permits charged solutions with structures different from the Maxwell case, as illustrated by the constructions of Ay\'on-Beato and Garc\'ia, Bronnikov, and Dymnikova~\cite{AyonBeato,Bronnikov,Dymnikova}. These examples provide theoretical context rather than implying that the truncated geometry studied here is regular. Effective-geometry and causality analyses further clarify why the electromagnetic characteristics require independent attention~\cite{Novello,Obukhov,Gibbons}. We adopt the magnetically charged Einstein-nonlinear-electrodynamics (ENED) background with an Euler-Heisenberg-type Lagrangian and polarization factors of the adopted model~\cite{dePaula}, retain their stated approximation order, and identify the admissible exterior domain before considering observables.

Quantum-electrodynamic corrections described by Euler-Heisenberg (EH) nonlinear electrodynamics modify the Reissner-Nordstr\"om (RN) geometry through a higher-order charge contribution proportional to $Q^{4}/r^{6}$. For the static Einstein-EH black hole, the optical sector has been studied in detail: Amaro and Mac\'ias analyzed the horizon structure, photon sphere, null geodesics, and shadow~\cite{ref1}, and later derived the exact gravitational lens equation with the associated lensing observables and time delays~\cite{ref2}, while the shadow radius and thermodynamic behavior of the charged solution were examined in Ref.~\cite{ref5}. Thermodynamics, critical behavior, and phase structure of EH-AdS black holes were investigated in Ref.~\cite{ref3}. The EH framework has also been extended to rotating configurations, for shadows and quasinormal modes~\cite{ref4}; to backgrounds with additional matter, such as a cloud of strings and dark matter~\cite{ref6}; to magnetically charged black holes with scalar hair, including particle motion, photon orbits, and Event Horizon Telescope constraints~\cite{ref7}; and to Einsteinian cubic gravity and AdS/dS backgrounds~\cite{ref8,ref9}. Static electrically charged, magnetically charged, and dyonic EH black holes were studied by Yajima and Tamaki~\cite{YajimaTamaki}. Most of these works address shadows, lensing, and thermodynamics. Vacuum birefringence in Einstein-nonlinear-electrodynamics black holes was considered by de~Paula et al., who showed that the two effective photon geometries produce distinct unstable light rings and two different shadows~\cite{dePaula}. The background spacetime considered here is equivalent to that of Ref.~\cite{dePaula} up to a rescaling of the nonlinear electrodynamics coupling parameter, while the effective photon geometries follow the corresponding polarization-dependent propagation laws.

The resulting distinction between spacetime and optical geometry becomes particularly important for strong-field photon propagation. Boillat's characteristic equations, effective null cones, and the factorization of the Fresnel equation establish how nonlinear electromagnetic interactions determine the propagation of each polarization~\cite{Novello,Obukhov,Gibbons,Boillat,DeLorenci}. These results motivate a separate radial potential and ray map for each polarization in the adopted model. A similar polarization dependence of photon propagation arises when the Maxwell field is coupled nonminimally to the Weyl tensor, where shadows have been studied for the two polarization modes and thin-disk emission has also been computed~\cite{WeylEM}. Black hole image formation also depends on emission and transfer: redshift and focusing were treated by Cunningham, and thin-disk images by Luminet~\cite{Cunningham,Luminet}. The distinction between the capture boundary, lensed emission, and successive photon subrings is developed in the analyses of Gralla, Holz, and Wald and of Johnson et al.~\cite{Gralla,Johnson}. We adapt the orbit-number and disk-transfer construction presented by Qi and Cai~\cite{QiCai} to both effective metrics. At fixed background parameters, identical static emissivity prescriptions in the two polarization modes, paired screen maps, and residual panels isolate their propagation differences. A brightness maximum also depends on the prescribed source and need not coincide with a geometric capture threshold.

Interferometric visibility provides a complementary description of these images by relating their brightness distribution to a signal measured as a function of baseline. Chen et al. developed an image-order and Fourier analysis for black holes with multiple photon spheres~\cite{ChenVisibility}.
We separate direct, lensed, and higher-order intensities for each mode, evaluate their signed visibility transforms, and compare their unresolved sum with a two-ring asymptotic model. Each polarization has its own potential maximum; matching their Fourier signatures to a two-photon-sphere calculation does not imply a shared inter-barrier trajectory. The comparison therefore tests how two independent optical geometries contribute to an unresolved interferometric signal.

Neutral massive particles probe the background spacetime through their orbital motion. When a stellar-mass compact object moves around a much heavier black hole, bound geodesics provide conservative reference trajectories for an extreme-mass-ratio inspiral; periodic members of this family offer a useful organization of its strong-field dynamics. The rational-frequency taxonomy of Levin and Perez-Giz, their separatrix analysis, and zoom-whirl studies organize the bound-orbit domain and its approach to plunge~\cite{Levin,Homoclinic,Glampedakis}. Fundamental-frequency and analytic-geodesic methods provide a broader framework for such classifications~\cite{Schmidt,Fujita}. Periodic motion has also been examined for black holes immersed in dark-matter halos~\cite{KingPO}. Approximate waveforms and fixed-orbit radiation snapshots have been developed through analytic, numerical, and augmented kludge constructions and black hole perturbation calculations~\cite{BarackCutler,Babak,Drasco,Chua}. Self-force theory describes the radiation reaction required for a complete evolving inspiral~\cite{BarackPound}. Here the background metric determines the massive-particle trajectories; photon effective metrics do not replace it. We calculate closed orbits, periods, and a physically admissible energy-angular-momentum domain with the energy scan restricted to values below unity, and use quadrupole snapshots to compare orbital timing rather than claim precision inspiral templates.

Accretion radiation connects this orbital dynamics to the energy released by matter approaching the black hole. The standard thin-disk framework was introduced by Shakura and Sunyaev and developed relativistically by Novikov and Thorne and Page and Thorne~\cite{Shakura,NovikovThorne,PageThorne}. Thorne's analysis distinguishes emitted energy from black hole evolution and photon capture~\cite{Thorne}. Relativistic transfer and image formation are described by Cunningham and Luminet, while multitemperature spectral models illustrate the roles of redshift, focusing, and the inner boundary~\cite{Cunningham,Luminet,Li,Zimmerman}.  Thin-disk flux, temperature, and spectral luminosity have also been computed for black holes in dark-matter halos~\cite{BoshkayevMNRAS,KurmanovDM,BoshkayevDM,HernquistDisk,KingDMDisk}. Non-LTE disk atmospheres and magnetic stresses provide examples of physical refinements beyond a local blackbody, zero-torque disk~\cite{DavisHubeny,AgolKrolik}. We derive the stationary orbital quantities, torque, flux, and temperature from the same spherical background, then distinguish emitted efficiency, direct face-on spectra, and ray-mapped isotropic-equivalent spectra. For each parameter choice, the same orbiting, opaque thermal disk is imaged through both photon geometries, allowing their polarization-dependent transfer to be compared at fixed intrinsic heating. This thermal source has a different velocity field, opacity, and emission law from the static prescribed emitters used for the earlier imaging and visibility calculations.

We extend the analysis of Ref.~\cite{dePaula} to interferometric visibility, neutral periodic orbits, and thin-disk radiation in the same background.
We study optical transfer, interferometric visibility, periodic-orbit classification, and stationary disk conservation laws in one background geometry with the same effective photon metrics. Sections~III-IV use prescribed static, optically thin emitters, whereas Section~VI uses an orbiting, optically thick thermal disk. These are separate source configurations, so their brightness and spectral results are not alternative observational predictions for one fixed accretion flow. A common source is imposed between polarization modes within each configuration; no common material configuration is imposed across these sections. Section~II introduces the background geometry, Section~III examines null geodesics and images, Section~IV develops interferometric visibility, Section~V analyzes periodic orbits and waveform snapshots, and Section~VI studies accretion radiation. The conclusion discusses the combined results and their limitations.

\section{Background spacetime in Einstein-nonlinear-electrodynamics}

We adopt the magnetically charged ENED black hole studied by de Paula et al.~\cite{dePaula}. Its nonlinear electromagnetic source has an Euler-Heisenberg-type Lagrangian. That paper calls this sample solution an Euler-Heisenberg (EH) black hole, while explicitly allowing a generic coupling rather than fixing the quantum-electrodynamic coefficient. We use ENED for the gravitational framework and EH-type for this particular matter model. The background spacetime metric $g_{\alpha\beta}$ governs neutral massive particles and material clocks; the polarization-dependent effective photon metrics $(\bar g_{\alpha\beta})_\pm$ govern electromagnetic characteristics. The latter describe two optical geometries associated with one black hole, not two distinct background spacetimes.

The background line element is
\begin{equation}
ds^2=-f(r)dt^2+f(r)^{-1}dr^2+r^2(d\theta^2+\sin^2\theta\,d\phi^2).\label{eq:geometry}
\end{equation}
For the magnetic EH-type solution of de Paula et al.~\cite{dePaula}, the lapse is
\begin{equation}
f(r)=1-\frac{2M}{r}+\frac{Q^2}{r^2}-\frac{2\mu Q^4}{5r^6}.\label{eq:lapse}
\end{equation}
We use $G=c=1$ unless physical units are restored explicitly. The parameters are $M$, magnetic charge $Q$, and EH-type NED coupling $\mu$, with $[\mu]=[M]^2$. Here $\mu$ is a generic free parameter of the Lagrangian; the scanned values are not assumed to be its fixed QED coefficient. The horizon is the largest positive root of $f(r)=0$. The metric reduces to RN when $\mu=0$, and to Schwarzschild when $Q=0$. The following sections use this same lapse throughout, including its circular-orbit and disk derivatives. The photon factors adopted here have the truncation order stated below.

The reference $Q/M=1$, $\mu=0$ is extremal RN, whereas $Q/M=1$ with positive $\mu$ has a simple, nonextremal outer horizon. With $x_h=r_h/M$, $\widehat\mu=\mu/M^2$, and $\epsilon=\sqrt{2\widehat\mu/5}$, the prescribed lapse gives
\begin{gather}
x_h^4(x_h-1)^2=\frac{2\widehat\mu}{5},\label{eq:horizonPolynomial}\\
\shortintertext{On the outer branch, the positive square root reduces the horizon condition to}
x_h^2(x_h-1)=\epsilon\quad(x_h>1),\label{eq:horizonCubic}\\
\shortintertext{Expanding this root in $\epsilon$ gives}
x_h=1+\epsilon-2\epsilon^2+O(\epsilon^3).\label{eq:horizonExpansion}
\end{gather}
Thus the leading horizon displacement is proportional to $\sqrt{\widehat\mu}$, rather than $\widehat\mu$. The exact root of this cubic within the adopted lapse is $x_h=1.077096671$ at $\widehat\mu=0.02$. Its surface gravity satisfies $M\kappa_h=(x_h-1)(3x_h-2)/x_h^3>0$. A regular Taylor expansion of the horizon in $\widehat\mu$ at fixed $Q/M<1$ is therefore not uniform as $Q/M\to1$. Horizon-sensitive source prescriptions require this distinction. The outer RN photon orbit at $r=2M$ remains nondegenerate, allowing the regular exterior critical-impact-parameter expansion developed in Section~IV.

\section{Null geodesics in the two effective photon metrics}

In nonlinear electrodynamics, electromagnetic rays need not follow the null cones of the background spacetime. For the magnetically charged EH-type ENED model considered by de Paula et al.~\cite{dePaula}, the two photon polarizations propagate in different effective geometries. Their background lapse agrees with Eq.~\eqref{eq:lapse}. We therefore adopt its magnetic-charge interpretation for the following optical analysis; an electric configuration requires its own effective-metric derivation.

The adopted two-invariant EH-type Lagrangian~\cite{dePaula} is
\begin{gather}
\mathcal L(F,G)=F-\mu\left(F^2+\frac74G^2\right),\label{eq:existing13}\\
\shortintertext{The first electromagnetic invariant entering this Lagrangian is}
F=F_{\alpha\beta}F^{\alpha\beta},\label{eq:invariantF}\\
\shortintertext{The second invariant contains the Hodge dual of the field tensor:}
G=F_{\alpha\beta}\,{}^\star F^{\alpha\beta}.\label{eq:invariantG}
\end{gather}
For the purely magnetic background, $F_{\theta\phi}=Q\sin\theta$, $F=2Q^2/r^4$ and $G=0$. The dependence on the second invariant still contributes to the characteristic equations of perturbations and produces the two polarization sectors.

The two effective line elements are
\begin{equation}
\begin{aligned}
d\bar s_\pm^2&=\frac{1}{G_1^\pm(r)}\left[-f(r)dt^2+\frac{dr^2}{f(r)}\right]\\
&\quad+\frac{r^2}{G_2^\pm(r)}\left(d\theta^2+\sin^2\theta\,d\phi^2\right).
\end{aligned}\label{eq:existing14}
\end{equation}
The magnetic factors of the adopted model~\cite{dePaula} are
\begin{gather}
G_1^+(r)=1+\frac{10\mu Q^2}{r^4},\label{eq:existing15}\\
\shortintertext{The angular factor for the same plus-polarization branch is}
G_2^+(r)=1-\frac{4\mu Q^2}{r^4},\label{eq:photonG2Plus}\\
\shortintertext{For the minus-polarization branch, the temporal and radial factor becomes}
G_1^-(r)=1-\frac{4\mu Q^2}{r^4},\label{eq:photonG1Minus}\\
\shortintertext{Its angular factor is instead}
G_2^-(r)=1-\frac{12\mu Q^2}{r^4}.\label{eq:photonG2Minus}
\end{gather}
The signs label the two polarization branches of the effective photon metrics. Both geometries recover the background metric for $\mu=0$ or $Q=0$. The effective inverse metrics retain terms through first order in $\mu$ and discard $O(\mu^2)$. Their inversion yields the displayed rational factors, but these expressions should not be interpreted as establishing higher-order accuracy of the effective-NED approximation. A first-order expansion gives $G_2^+/G_1^+=1-14\mu Q^2/r^4+O(\mu^2)$ and $G_2^-/G_1^-=1-8\mu Q^2/r^4+O(\mu^2)$.

Lorentzian signature requires that the optical domain have $f>0$, $G_1^\pm>0$, and $G_2^\pm>0$. For $\mu\geq0$, the plus-sector signature radius is $r_{\rm sig}^+=(4\mu Q^2)^{1/4}$, and the minus-sector radius is $r_{\rm sig}^-=(12\mu Q^2)^{1/4}$. To use both polarizations throughout the exterior, one requires $r_h>r_{\rm sig}^-$. For the parameter choices $\widehat\mu=0.02$ and $\widehat Q=0.3,0.7,1.0$, the corresponding $r_{\rm sig}^-/M$ values are approximately $0.3833$, $0.5857$, and $0.6999$, each below its horizon in Table~\ref{tab:exterior_boundaries}. This signature check does not by itself establish the accuracy of the NED truncation.

Restricting to the equatorial plane, let $\lambda_\pm$ be an affine parameter for the chosen effective metric. The conserved photon energy and angular momentum are $\mathcal E_\pm=-p_t$ and $\mathcal J_\pm=p_\phi$. Using the effective-geodesic description~\cite{dePaula}, the photon Hamiltonian and its null constraint are
\begin{equation}
\mathcal H_\pm=\frac12\left[-\frac{G_1^\pm}{f}\mathcal E_\pm^2
 +fG_1^\pm p_r^2+\frac{G_2^\pm}{r^2}\mathcal J_\pm^2\right]=0.\label{eq:existing16}
\end{equation}
Hamilton's equations consequently give
\begin{gather}
\frac{dt}{d\lambda_\pm}=\frac{G_1^\pm\mathcal E_\pm}{f},\label{eq:existing17}\\
\shortintertext{The conserved angular momentum similarly fixes the azimuthal evolution:}
\frac{d\phi}{d\lambda_\pm}=\frac{G_2^\pm\mathcal J_\pm}{r^2},\label{eq:photonPhi}\\
\shortintertext{The null constraint then determines the radial motion:}
\left(\frac{dr}{d\lambda_\pm}\right)^2=(G_1^\pm)^2\left[\mathcal E_\pm^2-\mathcal J_\pm^2U_\pm(r)\right],\label{eq:photonRadial}\\
\shortintertext{The polarization-dependent barrier is encoded in the reduced optical potential}
U_\pm(r)=\frac{f(r)G_2^\pm(r)}{r^2G_1^\pm(r)}.\label{eq:photonPotential}
\end{gather}
The photon constants are distinguished from the specific $E,L$ of a massive secondary. The optical potential is $\mathcal J_\pm^2U_\pm$, and its extrema determine the polarization-dependent light rings.

For the effective metrics of the adopted model, a circular null orbit at $r=r_{\rm LR}^\pm$ within the regular optical domain satisfies
\begin{gather}
U_\pm^\prime(r_{\rm LR}^\pm)=0,\label{eq:existing18}\\
\shortintertext{In terms of the adopted magnetic factors, this stationarity condition reads}
rf^\prime+rf\left(\frac{(G_2^\pm)^\prime}{G_2^\pm}-\frac{(G_1^\pm)^\prime}{G_1^\pm}\right)-2f=0.\label{eq:lightRingCondition}
\end{gather}
The second expression is evaluated at the light-ring radius. A local maximum, $U_\pm''<0$, is unstable, whereas a local minimum is stable. Thus the usual background condition $rf'-2f=0$ is generally insufficient to determine photon rings in this NED model. The two effective geometries can yield distinct light-ring radii even though they share the same $f(r)$.

With $b_\pm=\mathcal J_\pm/\mathcal E_\pm$, the critical impact parameter of an unstable light ring in the adopted effective metrics is
\begin{equation}
b_{\rm crit}^\pm=\frac{1}{\sqrt{U_\pm(r_{\rm LR}^\pm)}}
 =r_{\rm LR}^\pm\sqrt{\frac{G_1^\pm(r_{\rm LR}^\pm)}{f(r_{\rm LR}^\pm)G_2^\pm(r_{\rm LR}^\pm)}}.\label{eq:existing19}
\end{equation}
For an observer at infinity, where the magnetic factors approach unity, the relevant capture threshold sets the corresponding shadow radius in impact-parameter units. If several extrema occur, the capture boundary must be determined from the full optical potential rather than assigning a shadow edge to every light ring. For a ray with turning radius $r_0$, the turning condition is $b_\pm^2 U_\pm(r_0)=1$, and the trajectory obeys $|d\phi/dr|=b_\pm G_2^\pm/[r^2G_1^\pm\sqrt{1-b_\pm^2U_\pm}]$.

This polarization dependence is the vacuum birefringence associated with the two polarization branches. It does not replace the background metric in the neutral timelike geodesics below. The periodic-orbit and quadrupole-waveform calculations therefore continue to use Eq.~\eqref{eq:geometry}, while photon propagation uses the two effective geometries of Eq.~\eqref{eq:existing14}.

We next construct orbit numbers, transfer functions and thin-disk images for both effective metrics. We use the ENED lapse of Eq.~\eqref{eq:lapse} and retain the two effective photon metrics throughout. The adopted idealization is an infinitesimally thin, optically thin equatorial emitter viewed by an observer at infinity on the symmetry axis. Its material is static in the background coordinates, with $u_{\rm stat}^\alpha=f^{-1/2}(1,0,0,0)$ for $r>r_h$. Static support is imposed as a kinematic assumption; its force balance and mass transport are not calculated. These prescribed sources are distinct from the circularly orbiting thermal disk of Section~VI. A photon's orbital plane contains that axis, so spherical symmetry reduces ray tracing to a two-dimensional plane. The azimuthal variable below measures the angle within that ray plane, which can be rotated into the equatorial plane to calculate the trajectory.

Set $u=1/r$, and write $w_\pm(u)=G_2^\pm(1/u)/G_1^\pm(1/u)$. Dividing the radial equation by the square of the angular equation in Eqs.~\eqref{eq:photonPhi}-\eqref{eq:photonRadial}, then using $du/dr=-1/r^2$, yields the inverse-radius orbit equation:
\begin{equation}
\left(\frac{du}{d\phi}\right)^2=\mathcal D_\pm(u;b)
 =\frac{1}{w_\pm(u)^2}\left[\frac1{b^2}-u^2f(1/u)w_\pm(u)\right].\label{eq:existing20}
\end{equation}
The polarization label on $b$ is suppressed when an individual sector is understood. Unlike the background expression, both the potential term and the prefactor depend on polarization. Explicitly, with $s=\mu Q^2u^4$, one has $w_+=(1-4s)/(1+10s)$ and $w_-=(1-12s)/(1-4s)$. Define $\mathcal U_\pm(u)=u^2f(1/u)w_\pm(u)$; this is $U_\pm(r)$ in inverse-radius coordinates.

Let $u_h=1/r_h$. For a ray traced from infinity, $b<b_{\rm crit}^\pm$ leads to capture, and $b>b_{\rm crit}^\pm$ gives an exterior turning point on the incoming branch.
The accumulated ray-plane angle is
\begin{equation}
\Phi_\pm(b)=
\begin{cases}
\displaystyle\int_0^{u_h}\frac{w_\pm(u)\,du}{\sqrt{b^{-2}-\mathcal U_\pm(u)}},&b<b_{\rm crit}^\pm,\\[6pt]
\displaystyle2\int_0^{u_t}\frac{w_\pm(u)\,du}{\sqrt{b^{-2}-\mathcal U_\pm(u)}},&b>b_{\rm crit}^\pm.
\end{cases}\label{eq:existing21}
\end{equation}
\begin{figure*}[!tp]
	\centering
	\includegraphics[width=.94\linewidth, keepaspectratio]{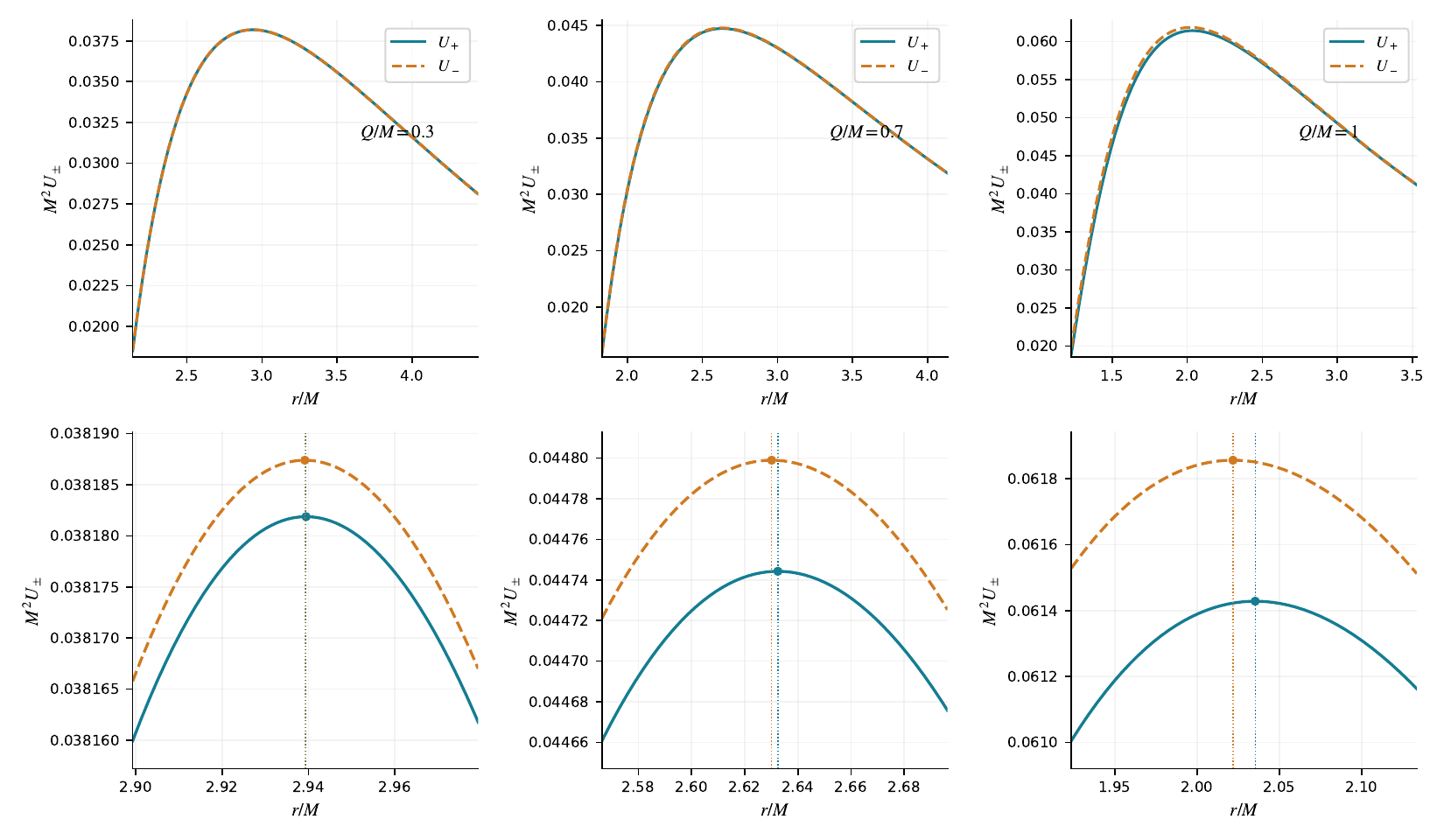}
	\caption{Reduced effective null potentials $M^2U_\pm$ for $\widehat\mu=0.02$, with $\widehat Q=0.3,0.7,1.0$ from left to right. Top row: restricted radial ranges around the potential barriers. Bottom row: separately scaled magnifications of the maxima, resolving the small polarization splitting. Solid blue and dashed orange curves show the plus and minus metrics, respectively. Dots mark the maxima, and vertical dotted lines identify their radii. Both axes in every panel remain linear; the lower panels use narrower radial and potential intervals. Numerical peak values are given in Table~\ref{tab:optical_maxima}.}
	\label{fig:optical6}
\end{figure*}

Here $u_t$ is the smallest positive root of $b^{-2}=\mathcal U_\pm(u)$, reached before the potential maximum. A scattering ray has deflection angle $\alpha_\pm=\Phi_\pm-\pi$. At the critical impact parameter the limiting orbit approaches the unstable light ring and $\Phi_\pm$ diverges.

Following the orbit-number classification of thin-disk images~\cite{Gralla}, define
\begin{gather}
n_\pm(b)=\frac{\Phi_\pm(b)}{2\pi},\label{eq:existing22}\\
\shortintertext{The direct sector contains at most one equatorial crossing:}
\begin{aligned}n_\pm<3/4:\quad&\text{direct sector, at most}\\&\text{one disk crossing},\end{aligned}\label{eq:directClass}\\
\shortintertext{The intermediate winding interval defines the lensed sector:}
3/4<n_\pm<5/4:\quad\text{lensed sector, two crossings},\label{eq:lensedClass}\\
\shortintertext{Larger winding numbers define the photon-ring sector~\cite{Gralla}:}
\begin{aligned}n_\pm>5/4:\quad&\text{photon-ring sector, at least}\\&\text{three crossings}.\end{aligned}\label{eq:photonClass}
\end{gather}
These classes describe ray geometry, not necessarily nonzero disk emission. A crossing within a chosen emission-free inner edge contributes no intensity. Some captured direct-sector rays terminate before their first crossing. The equalities at the classification boundaries are limiting cases.

\begin{figure*}[!tp]\centering\includegraphics[width=.94\linewidth,height=.78\textheight,keepaspectratio]{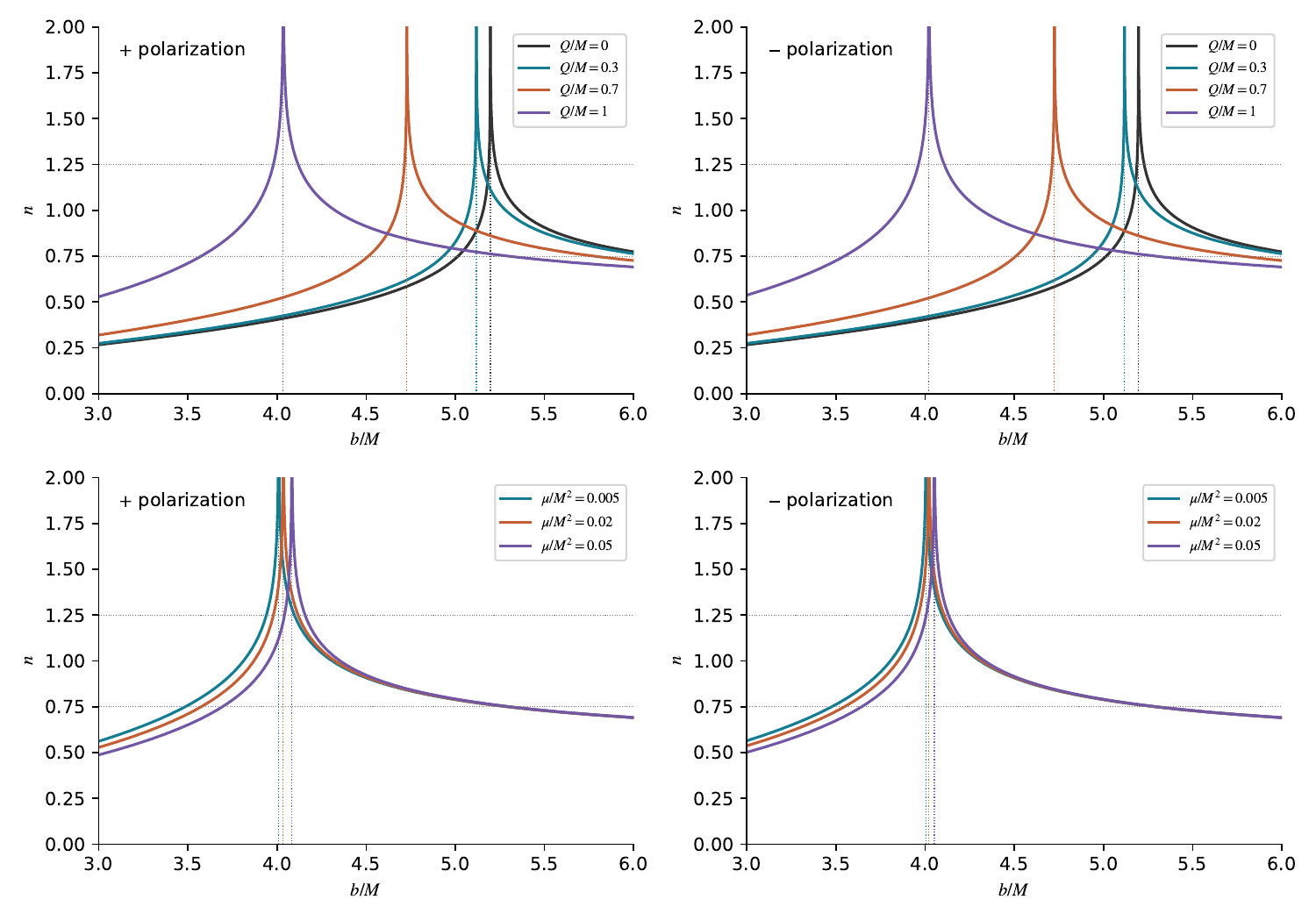}
\caption{Orbit numbers for both effective photon metrics. Left and right columns show the plus and minus modes. Top row: $Q/M=0,0.3,0.7,1.0$ at $\widehat\mu=0.02$, including the common Schwarzschild limit in black. Bottom row: $\widehat\mu=0.005,0.02,0.05$ at $Q/M=1$. Each vertical dotted line marks its curve's critical impact parameter. Horizontal lines at $n=3/4$ and $5/4$ distinguish direct, lensed and photon-ring sectors. The $Q=0$ curves coincide between the two modes and have $b_{\rm crit}/M=\sqrt{27}$.}\end{figure*}

Since the observer lies on the polar axis, the first equatorial-plane crossing occurs after a ray-plane angle $\pi/2$. Subsequent crossings are spaced by $\pi$. Accordingly, the transfer functions are defined by
\begin{gather}
\phi_m=\left(m-\frac12\right)\pi,\label{eq:existing23}\\
\shortintertext{Inverting the ray trajectory at each crossing angle defines its transfer radius:}
r_m^\pm(b)=\frac{1}{u_\pm(\phi_m;b)},\qquad m=1,2,3,\ldots,\label{eq:transferRadius}
\end{gather}
provided $\phi_m<\Phi_\pm(b)$. If this condition fails, the branch is absent rather than assigned zero radius. On the incoming segment, the cumulative integral in Eq.~\eqref{eq:existing21} is inverted to obtain $u_\pm(\phi;b)$. For a scattering ray, the outward segment is recovered by $u_\pm(\phi;b)=u_\pm(\Phi_\pm-\phi;b)$. This reflection prescription correctly accounts for the second and third disk intersections beyond the turning point.

The first branch generally maps disk radii across a broad impact-parameter interval. Higher branches appear near the critical impact parameter and can have steep slopes. Their radial image width follows $\delta b\simeq\delta r/|dr_m/db|$: a steep transfer function demagnifies the image of a narrow emitting annulus. This geometric compression explains why high-order images can be thin, but it does not prove that their intensity is negligible for every emissivity model.

\begin{figure*}[!tp]\centering\includegraphics[width=.94\linewidth,height=.76\textheight,keepaspectratio]{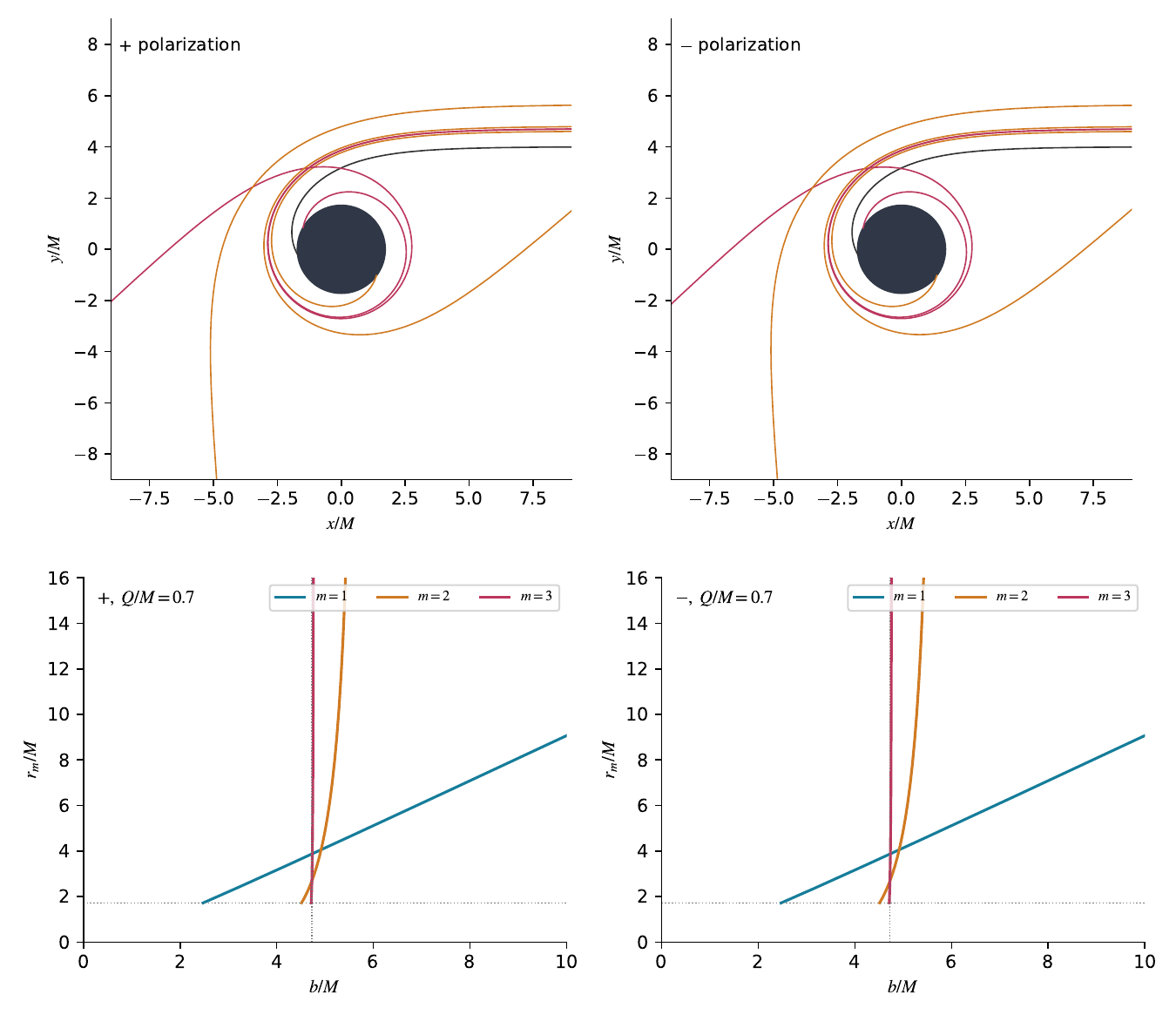}
\caption{Representative ray geometry and disk transfer at $Q/M=0.7$, $\widehat\mu=0.02$. Left and right columns show the plus and minus modes. Upper row: six trajectories with $b/b_{\rm crit}=0.85,0.98,0.999,1.001,1.02,1.2$, using each mode's own capture threshold; black, orange and red distinguish direct, lensed and photon-ring sectors. The filled disk marks the background horizon. Lower row: the first three transfer functions $r_m(b)$, in blue, orange and red for $m=1,2,3$. The vertical dotted line marks each mode's $b_{\rm crit}$ and the horizontal line the horizon. Broad first-crossing branches trace direct emission; steeper, narrower higher branches compress the images of emitting annuli. Trajectories and transfer functions show the path geometry and the corresponding emission-radius mapping in one comparison.}\end{figure*}

For the intensity calculation, we sum the redshifted emission over the intersections supplied by each polarization transfer function, using the usual intensity-transfer prescription~\cite{Cunningham,Luminet}:
\begin{gather}
I_\pm^{\rm obs}(b)=\sum_{m\geq1}g\bigl(r_m^\pm(b)\bigr)^4I_\pm^{\rm em}\bigl(r_m^\pm(b)\bigr),\label{eq:existing24}\\
\shortintertext{For the prescribed static emitters, the frequency shift is}
g(r)=\sqrt{f(r)}.\label{eq:staticRedshift}
\end{gather}
The background lapse enters this redshift factor because the emitter frequency is measured with the specified static material four-velocity. With $u_{\rm em}^\alpha=u_{\rm stat}^\alpha$ and an observer at infinity, the covector energy $\mathcal E=-p_t$ gives
\begin{gather}
\nu_{\rm em}\propto-p_\alpha u_{\rm em}^\alpha=\frac{\mathcal E}{\sqrt{f(r)}},\label{eq:existing25}\\
\shortintertext{At infinity, the same conserved photon energy determines the observed frequency:}
\nu_{\rm obs}\propto\mathcal E,\label{eq:observerFrequency}\\
\shortintertext{The resulting static-emitter redshift ratio is therefore}
\frac{\nu_{\rm obs}}{\nu_{\rm em}}=\sqrt{f(r)}.\label{eq:staticFrequencyRatio}
\end{gather}
Thus $f/G_1^\pm$ is not substituted for $f$ in material-clock redshift. If the ordinary geometric-optics invariant $I_\nu/\nu^3$ is adopted, the specific intensity has a factor $g^3$ and integration over observed frequency adds one more $g$, giving the bolometric $g^4=f^2$ prescription. A complete derivation of mode amplitudes in NED can require additional transport and refractive-medium factors; Eq.~\eqref{eq:existing24} is an adopted imaging approximation rather than a calculation of those factors. The imposed static velocity field supplies no orbital Doppler factor. The circular-emitter redshift of Section~VI therefore belongs to a different source model and is not substituted here. Absorption, mode conversion and plasma are omitted.

The source is taken to be unpolarized, with $I_+^{\rm em}=I_-^{\rm em}=I^{\rm em}/2$. In the plots, each polarization is normalized by its own emitted normalization $I_{0,\pm}=I_0/2$, so $I_\pm^{\rm obs}/I_{0,\pm}$ can be compared directly without an artificial factor of two. The unresolved total signal is the sum of the physical intensities of the two modes.

Use $x=r/M$, $x_h=r_h/M$, and the background timelike ISCO (innermost stable circular orbit) $x_{\rm ISCO}$. Let $x_{\rm ph}^{\rm bg}$ denote the unstable null circular radius of the background geometry, obtained from $rf'-2f=0$. We consider three prescribed static-emitter models, each with an emissivity profile motivated by idealized imaging studies~\cite{Gralla,QiCai,Bromley,Jaroszynski,ZengQuintessence,PengQuantum,ZengRegular,GuoHayward,WangHorndeski}:
\begin{gather}
\frac{I_1^{\rm em}(x)}{I_0}=\frac{\Theta(x-x_{\rm ISCO})}{[x-(x_{\rm ISCO}-1)]^2},\label{eq:existing26}\\
\shortintertext{Moving the common emitting edge to the background photon radius defines model 2:}
\frac{I_2^{\rm em}(x)}{I_0}=\frac{\Theta(x-x_{\rm ph}^{\rm bg})}{[x-(x_{\rm ph}^{\rm bg}-1)]^3},\label{eq:sourceModel2}\\
\shortintertext{Model 3 instead uses the horizon in the support and normalization of its emissivity:}
\begin{aligned}\frac{I_3^{\rm em}(x)}{I_0}&=\frac{\pi/2-\arctan[x-(x_{\rm ISCO}-1)]}{\pi/2-\arctan[x_h-(x_{\rm ISCO}-1)]}\\&\quad\times\Theta(x-x_h).\end{aligned}\label{eq:sourceModel3}
\end{gather}
The Heaviside step function is defined to vanish at the inner edge. Model 1 starts outside the massive-particle ISCO; model 2 permits emission deeper into the strong-field region; model 3 extends to just outside the horizon and decays smoothly outward. In all three prescriptions the emitter remains static wherever it emits. In particular, model 3 does not describe an infalling or plunging flow near the horizon. These profiles are imposed emissivities, with no hydrodynamic solution or thermal energy balance, and the support required to maintain static material is outside the model. The ISCO in a prescribed radial profile sets a length scale or boundary; it does not assign circular motion to that source. The normalized radius $x$ makes the offsets and arctangent arguments dimensionless.

\begin{figure*}[!tp]\centering\includegraphics[width=.94\linewidth,height=.78\textheight,keepaspectratio]{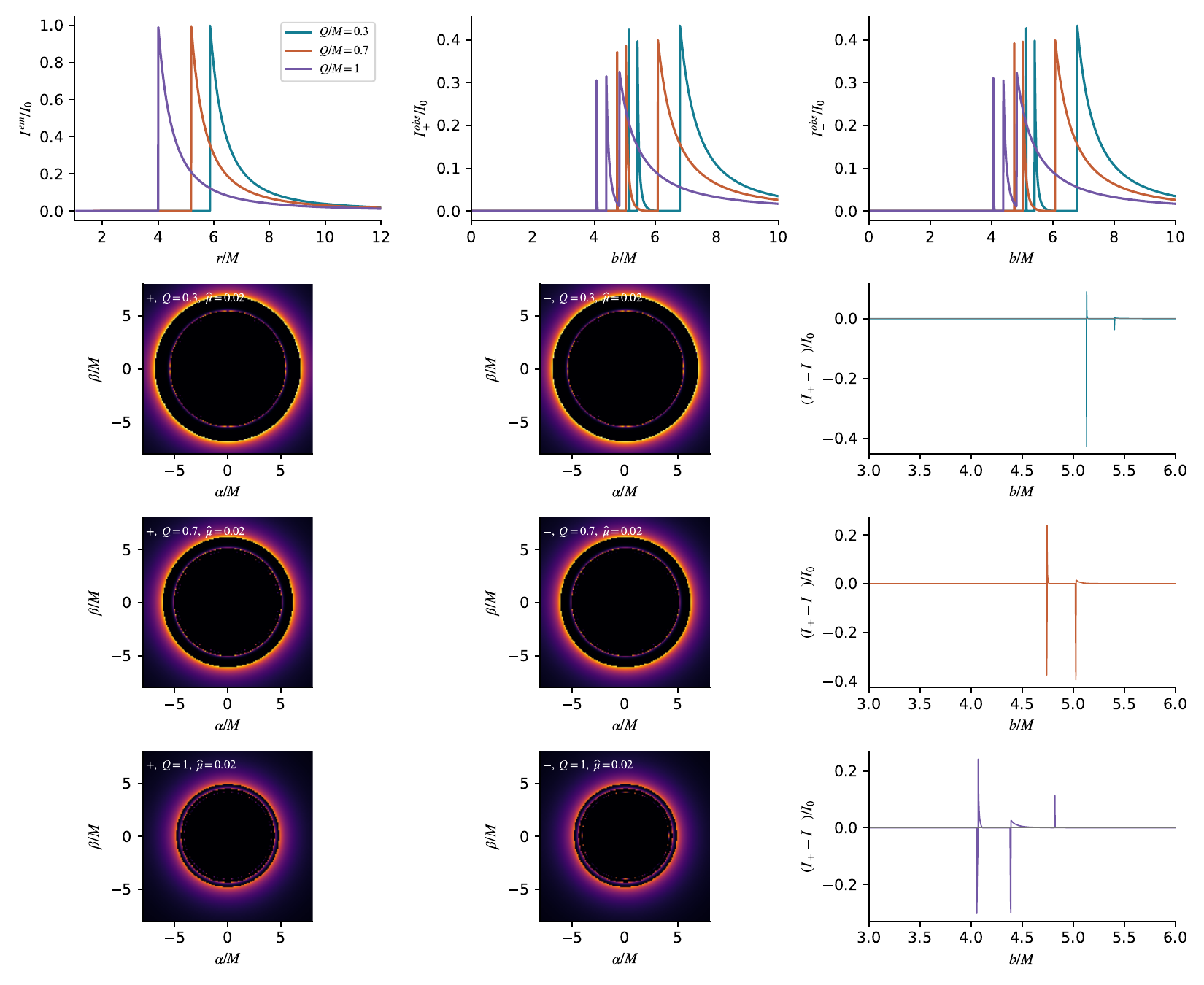}
\caption{Model 1: prescribed static, optically thin ISCO-edge inverse-square emissivity, for the charge scan. The emitter uses $g=\sqrt f$ and differs from the orbiting thermal disk of Section~VI. Top row: common emitted radial profile, observed plus-mode intensity, and observed minus-mode intensity. The three parameter values are $\widehat Q=0.3,0.7,1.0$ at $\widehat\mu=0.02$. Lower rows follow these values in ascending order: each row shows the plus-mode screen map, minus-mode screen map, and the difference of their radial observed intensities at the same $b$. All sums retain the first three disk crossings. The maps share one linear brightness scale within the figure, from black at zero to yellow-white at the maximum of the six radial intensity profiles; screen coordinates span $[-8M,8M]$. Intensity curves and maps are normalized by the per-mode emission amplitude $I_{0,\pm}=I_0/2$; the plotted shorthand $I_0$ denotes that per-mode normalization.}\end{figure*}

For model 2, the same $x_{\rm ph}^{\rm bg}$ is deliberately used for both modes. Replacing it separately by $x_{\rm LR}^+$ and $x_{\rm LR}^-$ would compare two different emitting disks as well as two different propagation laws. The latter choice is possible as a separate source model, but is not used for the polarization comparison here. Likewise the ISCO and horizon are those of the material background, not of either photon metric.

The matching of material sources between polarization modes holds at fixed $(Q,\mu)$. Across parameter choices the prescribed source family also changes because its inner edges and normalization depend on the background geometry. Model 3 is especially sensitive to the extremal Maxwell reference. At $Q/M=1$, $\mu=0$, one has $x_{\rm ISCO}=4$ and $A_0=\pi/2+\arctan2$. Since the ISCO displacement is $O(\widehat\mu)$, its normalized emissivity at any fixed $x>1$ has the expansion
\begin{equation}
\frac{I_3^{\rm em}(x;\widehat\mu)}{I_3^{\rm em}(x;0)}
=1+\frac{1}{5A_0}\sqrt{\frac{2\widehat\mu}{5}}+O(\widehat\mu).\label{eq:model3Normalization}
\end{equation}
The horizon therefore influences both the radial support and normalization of the prescribed emissivity. At $\widehat\mu=0.02$, the denominator change supplies a multiplicative factor $1.005925$ relative to RN; including the ISCO shift, the emitted intensity at $x=3$ increases by approximately $0.633\%$. Although $f^2$ suppresses emission immediately outside the horizon, it does not remove this normalization response at larger radii. Consequently, changes between different background parameters in these source-family images include source changes as well as geometric transfer. A separate transfer-only comparison would require one fixed radial emissivity, normalization, and inner edge outside all compared horizons. For example, $x_{\rm cut}=1.12$ is outside the horizons at $Q/M=1$ over $0\leq\widehat\mu\leq0.05$.

For a face-on observer at infinity, screen coordinates $(\alpha,\beta)$ satisfy
\begin{gather}
b=\sqrt{\alpha^2+\beta^2},\label{eq:existing27}\\
\shortintertext{The radial profile then determines the circularly symmetric screen intensity:}
\mathcal I_\pm(\alpha,\beta)=I_\pm^{\rm obs}\left(\sqrt{\alpha^2+\beta^2}\right).\label{eq:screenIntensity}
\end{gather}
These are circularly symmetric screen maps in impact-parameter units. The ray-plane coordinates used for trajectory figures, by contrast, are $x_{\rm ray}=r\cos\phi$ and $y_{\rm ray}=r\sin\phi$, and are not observer-screen coordinates or proper-distance embeddings.

The displayed intensities retain $m=1,2,3$ to include the first three transfer branches. The truncation is a stated imaging choice; higher intersections exist arbitrarily close to the light ring and are not excluded by the geometry. At each parameter choice, the two polarizations use the same static source profile, screen extent and brightness scale. This matching applies within that prescribed source configuration and does not identify it with the orbiting thermal disk of Section~VI. Narrow rings may be less visible in the finite-resolution image than in the resolved one-dimensional intensity curves.

\begin{figure*}[!tp]\centering\includegraphics[width=.94\linewidth,height=.78\textheight,keepaspectratio]{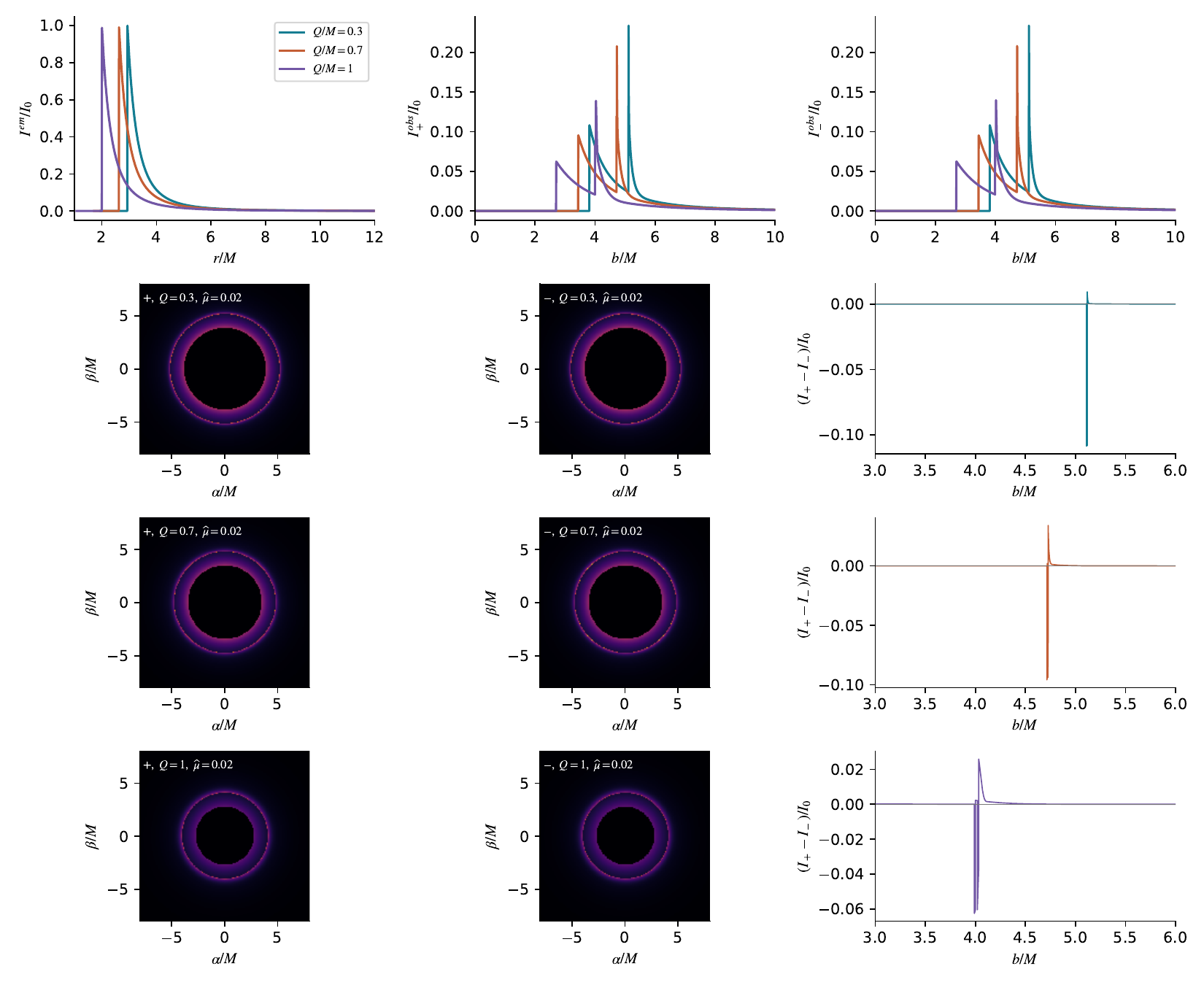}
\caption{Model 2: prescribed static, optically thin source with a common background-photon-radius edge and inverse-cubic emissivity, for the charge scan. Both modes use $g=\sqrt f$ for this source; the thermal disk of Section~VI is a separate configuration. Top row: common emitted radial profile, observed plus-mode intensity, and observed minus-mode intensity. The three parameter values are $\widehat Q=0.3,0.7,1.0$ at $\widehat\mu=0.02$. Lower rows follow these values in ascending order: each row shows the plus-mode screen map, minus-mode screen map, and the difference of their radial observed intensities at the same $b$. All sums retain the first three disk crossings. The maps share one linear brightness scale within the figure, from black at zero to yellow-white at the maximum of the six radial intensity profiles; screen coordinates span $[-8M,8M]$. Intensity curves and maps are normalized by the per-mode emission amplitude $I_{0,\pm}=I_0/2$; the plotted shorthand $I_0$ denotes that per-mode normalization.}\end{figure*}

To calculate each curve, we first find the largest positive root of $f=0$, check the exterior magnetic factors, and locate the maximum of $U_\pm$. The critical impact parameter then follows from Eq.~\eqref{eq:existing19}. Scattering turning points are bracketed on $0<u<u_{\rm LR}^\pm$, avoiding the inner roots that do not belong to the incident ray. For scattering integrations we use $u=u_t(1-z^2)$ to cancel the integrable square-root singularity. For captured rays the integration passes through the neighborhood of the potential maximum and ends at $u_h$. The transfer curves are obtained by monotone interpolation of the cumulative angle, with grid points concentrated near the critical impact parameter.

Independent adaptive quadrature and cumulative-angle integration agree within $1.5\times10^{-6}$ relative error for the sampled near-critical and ordinary rays; $Q=0$ recovers $b_{\rm crit}=3\sqrt3M$ in both modes. The parameter comparisons use $\widehat Q=0.3,0.7,1.0$ at $\widehat\mu=0.02$, and $\widehat\mu=0.005,0.02,0.05$ at $\widehat Q=1$. These choices have regular photon metrics throughout the exterior. Photon-ring boundaries are calculated independently for each mode by solving $n_\pm(b)=3/4$ and $n_\pm(b)=5/4$ on both sides of $b_{\rm crit}^\pm$; no common impact-parameter interval is imposed on the two polarizations.

The optical potentials constructed from the effective photon metrics of the adopted model approach zero both at the regular horizon and at infinity. For the displayed parameter choices each sector has one exterior maximum, representing an unstable light ring. Table~\ref{tab:optical_maxima} reports the maximum locations and heights calculated here at $\widehat\mu=0.02$. Figure~1 shows the corresponding profiles. The plus-sector maximum lies slightly farther out and is slightly lower for these examples. The splitting is small at low charge and becomes clearer near $Q/M=1$; this is a numerical result for the stated parameters, not a universal ordering for arbitrary NED theories.

\begin{table}[ht]\centering\small
\caption{Maxima of the reduced optical potentials at $\widehat\mu=0.02$. Heights are $M^2U_\pm^{\max}$, and $b_{\rm crit}/M=(M^2U^{\max})^{-1/2}$. Each maximum lies outside the horizon and has negative second radial derivative.}
\label{tab:optical_maxima}
\begin{tabular}{ccrrr}\toprule
$\widehat Q$ & polarization & $r_{\rm LR}/M$ & $M^2U^{\max}$ & $b_{\rm crit}/M$\\\midrule
0.3&$+$&2.939419&0.03818187&5.117660\\
0.3&$-$&2.939133&0.03818739&5.117289\\
0.7&$+$&2.632457&0.04474424&4.727499\\
0.7&$-$&2.630141&0.04479895&4.724611\\
1.0&$+$&2.035339&0.06142821&4.034745\\
1.0&$-$&2.021814&0.06185587&4.020773\\\bottomrule
\end{tabular}\end{table}
At $Q=0$, both magnetic factors equal unity and all optical equations reduce to Schwarzschild, independent of $\mu$. Figure~2 includes this common baseline alongside the charged orbit-number comparisons. Figure~3 combines representative ray trajectories and the first three transfer functions at $Q/M=0.7$. Figures~4-6 compare the three emissivity models at fixed $\widehat\mu=0.02$ while varying charge. The source profiles and paired image maps expose how the emitting edge and radial falloff affect the direct brightness and higher-order ring structure.

\begin{figure*}[!tp]\centering\includegraphics[width=.94\linewidth,height=.78\textheight,keepaspectratio]{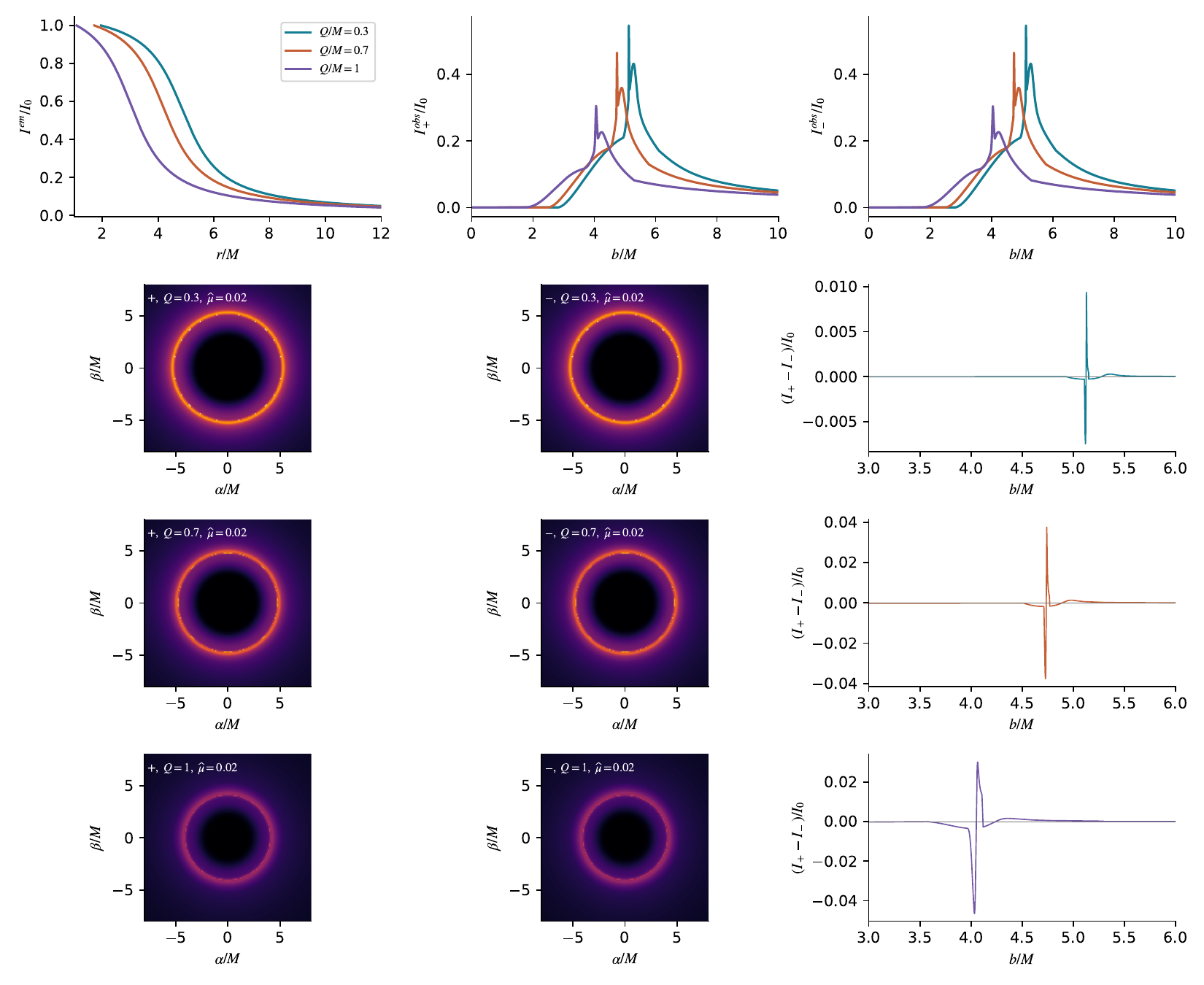}
\caption{Model 3: prescribed static, optically thin horizon-edge arctangent emissivity, for the charge scan. Material is held static at $r>r_h$ and uses $g=\sqrt f$; the prescription is not a plunging flow or the orbiting thermal disk of Section~VI. Top row: common emitted radial profile, observed plus-mode intensity, and observed minus-mode intensity. The three parameter values are $\widehat Q=0.3,0.7,1.0$ at $\widehat\mu=0.02$. Lower rows follow these values in ascending order: each row shows the plus-mode screen map, minus-mode screen map, and the difference of their radial observed intensities at the same $b$. All sums retain the first three disk crossings. The maps share one linear brightness scale within the figure, from black at zero to yellow-white at the maximum of the six radial intensity profiles; screen coordinates span $[-8M,8M]$. Intensity curves and maps are normalized by the per-mode emission amplitude $I_{0,\pm}=I_0/2$; the plotted shorthand $I_0$ denotes that per-mode normalization.}\end{figure*}

The numerical boundaries calculated here and reported in Table~\ref{tab:orbit_parameters} identify the direct, lensed and photon-ring sectors without relying on a broad visual color band. Denote the lower and upper roots of $n=3/4$ by $b_D^<$ and $b_D^>$, and those of $n=5/4$ by $b_P^<$ and $b_P^>$. The ordering is $b_D^<<b_P^<<b_{\rm crit}<b_P^><b_D^>$. Direct rays occupy $b<b_D^<$ or $b>b_D^>$; lensed rays occupy $(b_D^<,b_P^<)$ or $(b_P^>,b_D^>)$; photon-ring rays occupy $(b_P^<,b_{\rm crit})$ or $(b_{\rm crit},b_P^>)$. The critical ray itself has infinite limiting winding and is not a finite-time scattering or capture trajectory. Thus $b_{\rm crit}$ is the capture/scattering threshold fixed by the potential maximum, not a lensed-image boundary $b_{\rm lensed}$. The lensed sectors have two disjoint intervals bounded by $b_D^{<,>}$ and $b_P^{<,>}$; Table~\ref{tab:orbit_parameters} lists these boundaries explicitly.

\begin{table}[ht]
	\centering
	\small
	\caption{Orbit-number transition impact parameters at $\widehat\mu=0.02$, in units of $M$. All boundaries are solved separately for the two photon metrics. $b_D^{<,>}$ correspond to $n=3/4$, and $b_P^{<,>}$ to $n=5/4$.}
	\label{tab:orbit_parameters}
	\resizebox{\columnwidth}{!}{%
		\begin{tabular}{ccrrrrr}
			\toprule
			$\widehat Q$ & mode & $b_D^<$ & $b_P^<$ & $b_{\rm crit}$ & $b_P^>$ & $b_D^>$ \\
			\midrule
			0.3 & $+$ & 4.932426 & 5.108830 & 5.117660 & 5.150724 & 6.100011 \\
			0.3 & $-$ & 4.931995 & 5.108454 & 5.117289 & 5.150372 & 6.099819 \\
			0.7 & $+$ & 4.512387 & 4.714965 & 4.727499 & 4.769343 & 5.778349 \\
			0.7 & $-$ & 4.508867 & 4.711996 & 4.724611 & 4.766656 & 5.777036 \\
			1.0 & $+$ & 3.574527 & 3.979733 & 4.034745 & 4.117979 & 5.297523 \\
			1.0 & $-$ & 3.546759 & 3.962268 & 4.020773 & 4.106618 & 5.293628 \\
			\bottomrule
		\end{tabular}%
	}
\end{table}
Near an unstable light ring, $\mathcal U_\pm(u)$ has the local expansion $\mathcal U_{\max}+\tfrac12\mathcal U_{\pm}''(u_{\rm LR})(u-u_{\rm LR})^2$ with $\mathcal U_{\pm}''<0$. Substitution into Eq.~\eqref{eq:existing21} produces a logarithmic divergence $\Phi_\pm\sim-a_\pm\ln|b/b_{\rm crit}^\pm-1|+\text{constant}$, with $a_\pm=\sqrt{2}\,w_\pm(u_{\rm LR})/\sqrt{-\mathcal U_\pm''(u_{\rm LR})}$. This behavior explains the narrow higher-order transfer branches and why resolving points close to each polarization's own critical impact parameter is essential.

For the chosen charge scan, increasing $Q$ lowers the critical impact parameter, shifting the optical features inward in screen impact-parameter coordinates. At $Q=M$, increasing positive $\mu$ over the plotted interval shifts the critical impact parameters outward. The two modes do not share identical orbit-number boundaries, transfer functions or brightness peaks. The separation of their capture thresholds is not, however, a guarantee of two visibly resolved brightness rings: the source inner edge, relative branch intensities and finite resolution also matter.

Model 1 suppresses all intersections inside the background ISCO, so its broad direct image follows a comparatively distant source while higher-order features remain narrow. Model 2 moves the common emission edge inward and makes the photon-region structure more conspicuous. Model 3 prescribes static emission close to the horizon, with the $f^2$ factor strongly suppressing contributions from emitters approaching $r_h$. This is a transfer diagnostic for an imposed source, not a calculation of radiation from plunging accreting gas. A bright radial maximum should therefore not automatically be identified with $b_{\rm crit}$; it can arise from the emissivity edge mapped through any allowed transfer branch.

For each static-emitter configuration, the difference profiles displayed beside the two-mode images use the same impact parameter, comparing $I_+^{\rm obs}(b)-I_-^{\rm obs}(b)$ after interpolation onto a common grid. Their sign changes principally trace small displacements of intensity features rather than a universal ordering of the two polarization brightnesses. The following section applies the same static, optically thin transfer prescription to an explicitly specified tapered source for interferometric visibility, before the discussion of neutral massive-particle periodic orbits.

\section{Polarization-resolved images and interferometric visibility}

We extend the image and visibility calculations presented in Secs.~III-IV of Chen, Wang and Yang~\cite{ChenVisibility} to the two effective photon metrics~\cite{dePaula}. The characteristic formulation and birefringent interpretation follow Refs.~\cite{Novello,Obukhov,Gibbons,Boillat,DeLorenci}; the emission-transfer and photon-subring context is supplied by Refs.~\cite{Cunningham,Luminet,Gralla,Johnson}. The correspondence is between image contributions and Fourier signals, rather than between the radial mechanics of the two models. In the model of Chen et al., a single effective potential has two maxima. In the ENED examples, each polarization has one exterior maximum of its own potential. A photon stays in its polarization sector in our approximation and cannot reflect between a plus-sector barrier and a minus-sector barrier. The inter-barrier trajectories and associated broad photon-ring structure in their double-sphere model therefore do not arise simply by identifying our two light rings with theirs. The dependence of shadows, rings and optical brightness on the accretion prescription has also been examined for a magnetically charged regular black hole~\cite{ZareOptics}. This motivates keeping the source prescription common to both polarization sectors when comparing their image contributions and visibility signals.

For a face-on source, let $b$ be the impact parameter and let $\beta=b/D$ be its small angular coordinate, with $D$ the source distance. Here $D/M=100$ is used solely to express the angular axes and baseline scales. The rays are calculated for an observer at infinity. This represents a deliberate modification of the finite-radius observer configuration $r_o=100M$ used by Chen et al.; it is not an exact reconstruction of that finite-observer geometry. Our $\beta$ is unrelated to the inverse-radius variable $u=1/r$ used earlier. In this section $u$ instead denotes the interferometric baseline in wavelengths, following Ref.~\cite{ChenVisibility}; the distinct meanings are confined to their respective sections.

The transfer radius of the $n$th emitting crossing obeys
\begin{gather}
\phi_\pm(r_n^\pm;b)=\left(n-\frac12\right)\pi,\label{eq:visCross}\\
\shortintertext{The contribution from that crossing includes the static-emitter bolometric redshift:}
I_{n,\pm}(\beta)=f(r_n^\pm)^2I_{\rm em,\pm}(r_n^\pm).\label{eq:visCrossIntensity}
\end{gather}
Missing intersections contribute zero. For scattering rays the inward and outward branches are both included; captured rays terminate at the horizon. The material emitter in this section is static, with the same four-velocity assumption as in Section~III, so $g^4=f^2$ for an observer at infinity. The visibility below is consequently a Fourier diagnostic of a prescribed static source. It is not the visibility of the rotating thermal disk in Section~VI. The ordinary bolometric intensity-transport prescription already stated is used with the two ray maps of the adopted model. It does not introduce a new calculation of NED mode-amplitude transport.

Chen et al. take an optically thin emitter extending from the horizon outward with intensity proportional to $r^{-2}$. Such a source, extended to infinity without a field-of-view restriction, has a logarithmically divergent total flux. To define a finite zero-baseline visibility we use a smooth outer taper and an angular acceptance:
\begin{gather}
\begin{aligned}I_{\rm em,\pm}(r)&=\frac{I_0}{2}\left(\frac Mr\right)^2\exp\left[-\left(\frac r{30M}\right)^4\right],\\&\hspace{75pt}r>r_h,\end{aligned}\label{eq:visSource}\\
\shortintertext{The shared smooth angular acceptance is chosen as}
A(\beta)=\exp\left[-\left(\frac{D\beta}{12M}\right)^8\right].\label{eq:visAcceptance}
\end{gather}
The factor one half assigns equal emitted intensity to the two incoherent polarization modes. The taper and acceptance are shared by the modes; they suppress artificial oscillations from a sharp outer image edge while preserving the strong-field source profile. All images and visibilities below refer to the accepted intensity $\widetilde I_{n,\pm}=A I_{n,\pm}$. Equation~\eqref{eq:visSource} defines the static source specifically used for the visibility calculation; its taper and angular acceptance also distinguish it from the three illustrative profiles of Section~III. The source has no thermal energy-balance prescription and is physically distinct from the opaque, orbiting disk developed in Section~VI.

Following the image-order decomposition of Ref.~\cite{ChenVisibility}, define direct, lensed and photon-ring components by
\begin{gather}
\widetilde I_{\rm d,\pm}=\widetilde I_{1,\pm},\label{eq:visGroups}\\
\shortintertext{The second emitting crossing supplies the lensed contribution:}
\widetilde I_{\rm l,\pm}=\widetilde I_{2,\pm},\label{eq:visLensed}\\
\shortintertext{The retained higher crossings form the photon component:}
\widetilde I_{\rm ph,\pm}=\sum_{n=3}^8\widetilde I_{n,\pm},\label{eq:visPhoton}\\
\shortintertext{Adding these components gives the accepted intensity of one polarization:}
\widetilde I_\pm=\widetilde I_{\rm d,\pm}+\widetilde I_{\rm l,\pm}+\widetilde I_{\rm ph,\pm}.\label{eq:visTotal}
\end{gather}
This crossing-order classification differs from assigning a whole ray to the orbit-number intervals used earlier: one ray can collect emission at several crossings. The photon component is evaluated through eight intersections, rather than identified with the third intersection alone. Subrings of still higher order are omitted from the numerical image.

For a projected baseline vector $\boldsymbol u$, the complex visibility is the Fourier transform of the accepted sky brightness~\cite{Johnson,ChenVisibility}:
\begin{equation}
V_\pm(\boldsymbol u)=\int \widetilde I_\pm(\boldsymbol\beta)
 e^{-2\pi i\boldsymbol u\cdot\boldsymbol\beta}\,d^2\boldsymbol\beta
=2\pi\int_0^\infty\widetilde I_\pm(\beta)J_0(2\pi u\beta)\beta\,d\beta.\label{eq:visFT}
\end{equation}
Axisymmetry makes this visibility real, although it may change sign. The Hankel form follows by integrating the Fourier kernel around a circle. Each image-order component has the same transform, and its signed visibility is added before computing an amplitude. In terms of $\xi=b/M$ and $\kappa=uM/D$,
\begin{equation}
V_\pm(u)=2\pi\left(\frac MD\right)^2
\int_0^\infty\widetilde I_\pm(\xi)J_0(2\pi\kappa\xi)\xi\,d\xi.\label{eq:visScale}
\end{equation}
In particular, $V_\pm(0)$ is the accepted image flux. The direct and lensed transforms extend to $b=20M$, where the smooth acceptance is negligible. The photon transform additionally resolves the exponentially narrow neighborhoods of each mode's critical impact parameter. The figures show $Q/M=0.7,1.0$ at $\mu/M^2=0.02$, all within the admissible exterior optical domain. A small high-order contribution to the zero-baseline flux alone does not establish convergence at long baselines. We therefore test the signed photon visibility and its modulation minima throughout the numerical interval displayed in Figure~8.

\begin{figure*}[!tp]\centering\includegraphics[width=.94\linewidth,height=.78\textheight,keepaspectratio]{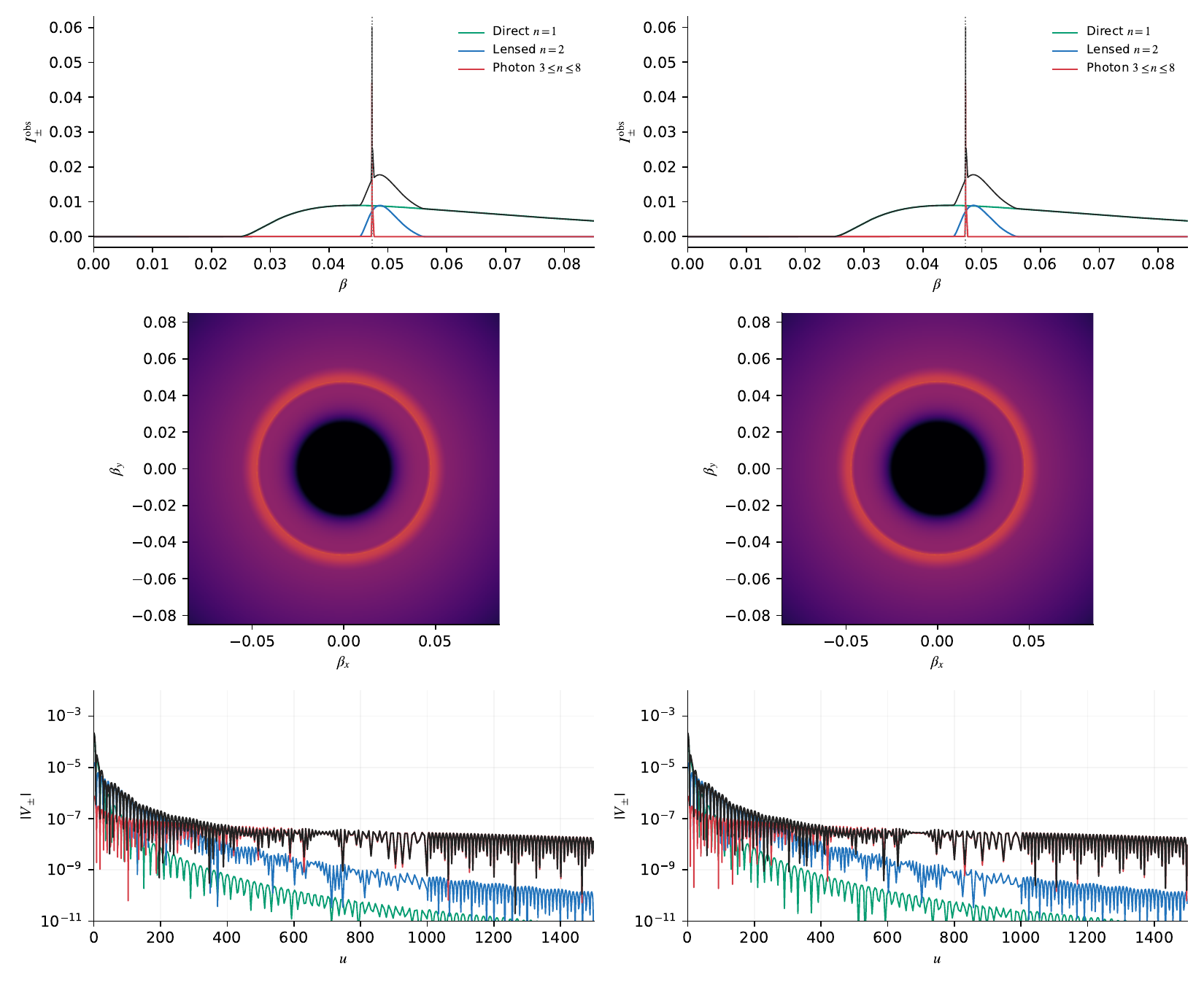}
\caption{Image and visibility decomposition for the prescribed static, optically thin source at $Q/M=0.7$ and $\mu/M^2=0.02$, with $g=\sqrt f$, calculated with the effective photon metrics using the approach of Ref.~\cite{ChenVisibility}. The orbiting thermal source of Section~VI is a different configuration. Left column: plus mode. Right column: minus mode. Top: direct, lensed and photon-crossing intensities in green, blue and red, with their total in black and critical position dotted. Middle: accepted face-on intensity maps with a square-root brightness stretch to expose faint structure; the modes share the same source prescription. Bottom: separate-mode visibility amplitudes with the same component colors and black total. The source and smooth angular acceptance are defined in Eqs.~\eqref{eq:visSource}-\eqref{eq:visAcceptance}; photon emission includes crossings three through eight. Angular coordinates use $D/M=100$ and infinity-observer ray maps. The baseline is in wavelengths.}\end{figure*}

For a receiver that sums the two polarization intensities, the unresolved Stokes-$I$ visibility is
\begin{gather}
\widetilde I_I=\widetilde I_++\widetilde I_-,\label{eq:visStokes}\\
\shortintertext{Linearity of the Fourier transform gives the corresponding visibility sum:}
V_I=V_++V_-,\label{eq:visStokesV}\\
\shortintertext{Division by the accepted zero-baseline flux defines the normalized signal:}
v_I(u)=\frac{V_I(u)}{V_I(0)}.\label{eq:visNormalized}
\end{gather}
In general $|V_I|\ne |V_+|+|V_-|$. Cancellation in the signed Fourier components is responsible for visibility nulls and can produce modulation. This is interference between spatial Fourier contributions to the brightness distribution, not coherence between electric fields in orthogonal polarization modes. The expression assumes the instrument measures their summed intensities; a polarization-resolved receiver instead measures the separate mode signals.

For the image-order test, define
\begin{gather}
V_{\rm ph}^{(N)}(u)=\sum_{\sigma=\pm}\sum_{n=3}^NV_{n,\sigma}(u),\label{eq:photonSum}\\
\shortintertext{For the long-baseline comparison, we retain the sign and rescale this sum as}
S_N(u)=\frac{\sqrt u\,V_{\rm ph}^{(N)}(u)}{F_{\rm ref}},\label{eq:signedConvergence}\\
\shortintertext{Its maximum-norm difference from the sixteen-order reference is measured by}
e_N=\frac{\max_{\mathcal B}|S_N-S_{16}|}{\max_{\mathcal B}|S_{16}|},\label{eq:visibilityError}
\end{gather}
where $\mathcal B=[3000,25000]$ and one fixed reference flux $F_{\rm ref}=V_I^{(8)}(0)$ is used for every $N$. This norm retains the signed visibility and avoids division by individual samples at a visibility null. We recalculate the photon sum through $N=8,10,12,14,16$, with logarithmic screen coordinates on both sides of each critical curve. A critical-point polynomial factorization removes cancellation in the optical potential, and hyperbolic substitutions regularize the angular integrals. Increasing the angular quadrature from 2049 to 4097 nodes and refining the composite Gaussian screen quadrature independently changes the reference $S_{16}$ by $1.67\times10^{-8}$ and $1.36\times10^{-7}$ of its interval maximum for $Q/M=0.7$ and $1$, respectively. The transforms are sampled on a unit baseline grid through $u=30000$. Table~\ref{tab:visibility_convergence} summarizes the signed errors and minimum shifts.

\begin{center}
\begin{minipage}{\columnwidth}
\captionsetup{hypcap=false}
\captionof{table}{Signed photon-visibility convergence on $\mathcal B=[3000,25000]$ at $\widehat\mu=0.02$. The errors $e_N$ are defined in Eq.~\eqref{eq:visibilityError}; the last column gives the largest modulation-minimum shift from $N=8$ to $16$.}
\label{tab:visibility_convergence}
\centering\small\setlength{\tabcolsep}{2.5pt}
\begin{tabular}{crrrr}
\toprule
$Q/M$ & $e_8$ & $e_{10}$ & $e_{12}$ & $\max|\delta u_{\min}|$\\\midrule
0.7 & $1.73\!\times\!10^{-7}$ & $5.06\!\times\!10^{-10}$ & $1.48\!\times\!10^{-12}$ & $<10^{-3}$\\
1.0 & $1.80\!\times\!10^{-5}$ & $1.86\!\times\!10^{-7}$ & $1.92\!\times\!10^{-9}$ & $0.017$\\\bottomrule
\end{tabular}
\end{minipage}
\end{center}

To identify the slow modulation minima reproducibly, introduce the quadrature envelope
\begin{gather}
\begin{aligned}Z_N(u)&=\frac{2\pi}{D^2}\sum_{\sigma=\pm}\int_0^\infty b\,\widetilde I_{{\rm ph},\sigma}^{(N)}(b)\\&\hspace{20pt}\times H_0^{(1)}\!\left(\frac{2\pi ub}{D}\right)db,\end{aligned}\label{eq:complexEnvelope}\\
\shortintertext{The modulus of this auxiliary transform defines the slow modulation envelope:}
\mathcal E_N(u)=\frac{\sqrt u}{F_{\rm ref}}|Z_N(u)|,\label{eq:quadratureEnvelope}\\
\shortintertext{The special function in the transform is the Hankel function of the first kind:}
H_0^{(1)}=J_0+iY_0.\label{eq:hankelDefinition}
\end{gather}
The real part of the corresponding scaled complex integral is exactly $S_N$; its modulus approaches the usual oscillation envelope in the narrow-ring limit. The dominant slow minima, found from $d\mathcal E_N^2/du=0$, are distinguished from the many fast zero crossings of the signed signal. At $N=16$ they occur at $u\simeq17351.60$ for $Q/M=0.7$, and at $u\simeq3616.26,10782.40,17917.96$ for $Q/M=1$ within $\mathcal B$. The two refined quadratures agree on these positions within $4\times10^{-4}$ in $u$. Extending $N=8$ to 16 shifts them by less than $10^{-3}$ and $0.017$, respectively. These tests support the eight-order photon approximation on the displayed numerical interval. Smaller order-increment differences do not establish a precision better than the independent quadrature check, and pointwise relative errors near nulls are not bounded by $e_N$. Figure~8 uses the refined photon transform; its separate ideal thin-ring panels do not extend this convergence claim to arbitrarily long baselines.

\begin{figure*}[!tp]\centering\includegraphics[width=.94\linewidth,height=.78\textheight,keepaspectratio]{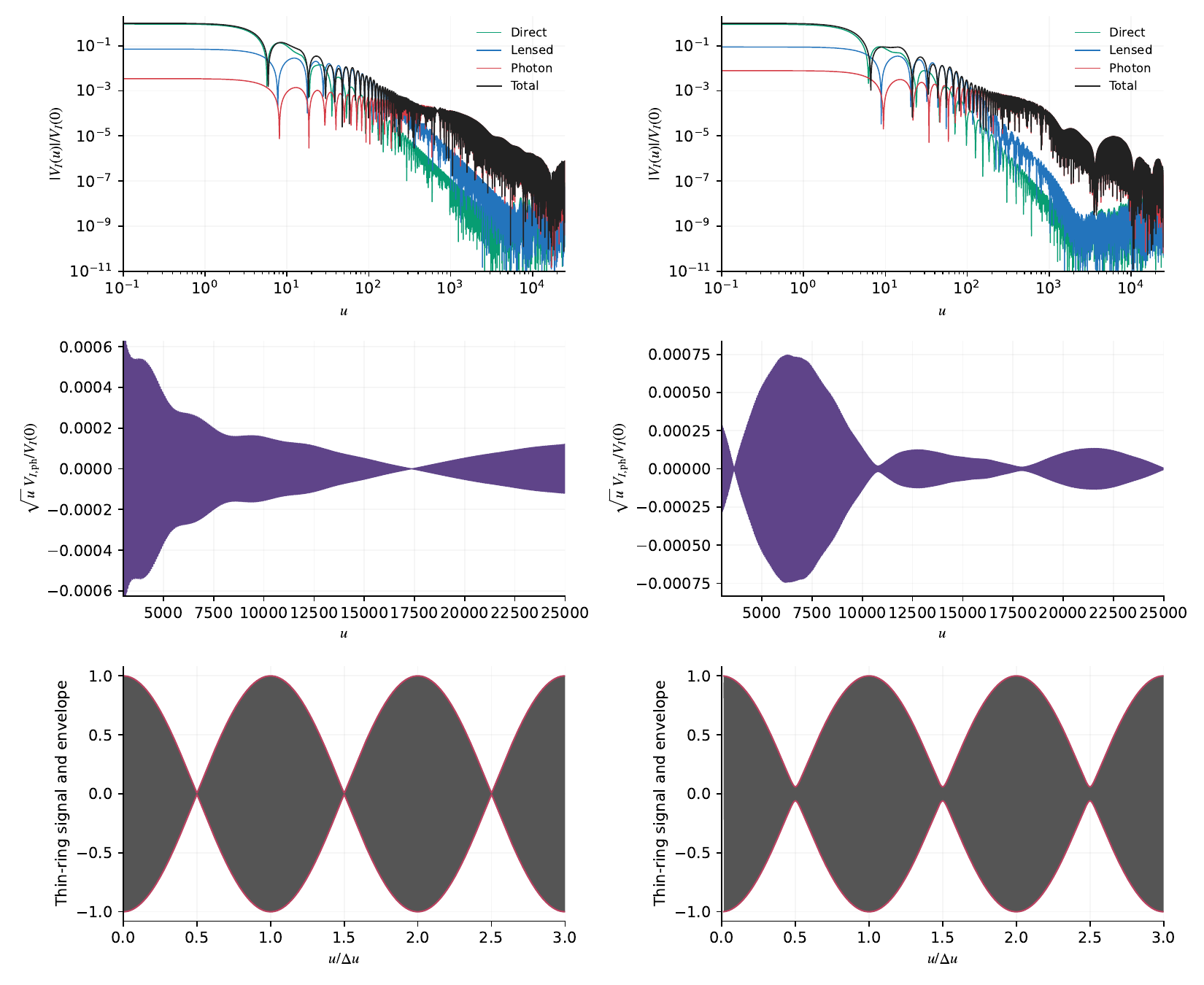}
\caption{Recomputed unresolved two-mode visibility and ideal thin-ring comparison for the prescribed static, optically thin source, with $g=\sqrt f$. Left: $Q/M=0.7$. Right: $Q/M=1$, both at $\mu/M^2=0.02$. Top: accepted Stokes-$I$ visibility amplitude and its direct, lensed and photon components, normalized by the fixed total zero-baseline flux. Signed contributions are added before taking the modulus. The photon transform uses refined screen quadrature and crossings three through eight, validated against sums through order 16. Middle: its signed finite-width photon visibility multiplied by $\sqrt u$ on $3000\leq u\leq25000$. The text reports signed-error norms and modulation-minimum convergence on this numerical interval. Bottom: ideal two-ring Bessel signals weighted by the recomputed accepted flux of crossings six through eight, with their asymptotic envelope in red. The axis uses $u/\Delta u_{\rm rat}$ with $\Delta u_{\rm rat}\simeq34626.34,7157.11$; the leading-order estimates are $34312.34,6666.67$. The ideal extension to three beat periods is not a finite-width numerical convergence claim. The orbiting thermal disk in Section~VI is a different source configuration. The two-mode thin-ring interpretation follows Ref.~\cite{ChenVisibility}.}\end{figure*}

To obtain the analytic counterpart of Chen et al.'s thin-ring calculation, approximate one sufficiently narrow subring in each sector by
\begin{gather}
\widetilde I_{\rm narrow,\pm}(\beta)=\frac{F_\pm}{2\pi\beta_\pm}\delta(\beta-\beta_\pm),\label{eq:visDelta}\\
\shortintertext{The angular radius of each ideal ring is set by its critical impact parameter:}
\beta_\pm=\frac{b_{\rm crit}^\pm}{D},\label{eq:angularCritical}\\
\shortintertext{The axisymmetric Fourier transform is the weighted Bessel sum~\cite{ChenVisibility}:}
V_{I,\rm narrow}(u)=F_+J_0(2\pi\beta_+u)+F_-J_0(2\pi\beta_-u).\label{eq:narrowVisibility}
\end{gather}
Here $F_\pm$ is the integrated accepted flux of the selected narrow subrings, not their peak intensity. The amplitudes in our analytic comparison are obtained from crossings six through eight. The delta-ring expression is an approximation to these selected components and is not substituted for the entire numerically ray-traced photon image.

The standard long-baseline thin-ring asymptotics~\cite{Johnson,ChenVisibility} give
\begin{gather}
\begin{aligned}\sqrt u\,V_{I,\rm narrow}&\simeq a_+\cos(2\pi\beta_+u-\pi/4)\\&\quad+a_-\cos(2\pi\beta_-u-\pi/4),\end{aligned}\label{eq:visAsym}\\
\shortintertext{The coefficient of each large-baseline oscillation is fixed by its ring flux:}
a_\pm=\frac{F_\pm}{\pi\sqrt{\beta_\pm}}.\label{eq:visAmplitude}
\end{gather}
Its envelope and beat spacing are
\begin{gather}
\begin{aligned}\mathcal A(u)=\Bigl\{a_+^2+a_-^2&+2a_+a_-\\&\times\cos[2\pi(\beta_+-\beta_-)u]\Bigr\}^{1/2},\end{aligned}\label{eq:visEnvelope}\\
\shortintertext{The spacing between successive slow minima follows from the difference of ring radii~\cite{ChenVisibility}:}
\Delta u=\frac{1}{|\beta_+-\beta_-|}=\frac{D}{|b_{\rm crit}^+-b_{\rm crit}^-|}.\label{eq:visBeat}
\end{gather}
For equal $a_\pm$, the oscillatory sum factors into a rapidly varying cosine and a slowly varying cosine; taking its amplitude gives minima spaced by $\Delta u$. Unequal mode weights keep the envelope minima above zero, with extrema $a_++a_-$ and $|a_+-a_-|$. Beat structure requires both contributions to remain appreciable and sufficiently narrow over the baselines considered. Finite subring widths, the smooth-source background, and omitted higher orders limit a literal thin-ring extrapolation. No detector sensitivity or angular-resolution forecast is made here.

Because Eq.~\eqref{eq:visBeat} contains the inverse of a small polarization splitting, its perturbative accuracy requires a separate check. Define the outer RN photon radius and impact parameter by
\begin{gather}
r_{\rm RN}=\frac{3M+\sqrt{9M^2-8Q^2}}2,\label{eq:rnPhotonRadius}\\
\shortintertext{The corresponding Maxwell lapse is}
f_{\rm RN}(r)=1-\frac{2M}{r}+\frac{Q^2}{r^2},\label{eq:rnLapse}\\
\shortintertext{Evaluating the capture threshold at the outer RN photon orbit gives}
b_{\rm RN}=\frac{r_{\rm RN}}{\sqrt{f_{\rm RN}(r_{\rm RN})}}.\label{eq:rnCritical}
\end{gather}
The first-order displacement of the light-ring radius does not contribute to the first-order value of the optical-potential maximum, because the RN potential is stationary at $r_{\rm RN}$. Expanding both the lapse and the optical factors consistently gives
\begin{gather}
\begin{aligned}b_{\rm crit}^\pm=b_{\rm RN}\Biggl[1+&\frac{\mu Q^4}{5r_{\rm RN}^6f_{\rm RN}(r_{\rm RN})}\\{}+&C_\pm\frac{\mu Q^2}{r_{\rm RN}^4}\Biggr]+O(M\widehat\mu^2),\end{aligned}\label{eq:criticalExpansion}\\
\shortintertext{Expansion of the plus-mode photon factors fixes the coefficient}
C_+=7,\label{eq:criticalCPlus}\\
\shortintertext{The minus-mode factors give the different coefficient}
C_-=4.\label{eq:criticalCMinus}
\end{gather}
The common lapse term cancels in the difference, leaving
\begin{gather}
b_{\rm crit}^+-b_{\rm crit}^-=\frac{3\mu Q^2b_{\rm RN}}{r_{\rm RN}^4}+O(M\widehat\mu^2),\label{eq:criticalSplitting}\\
\shortintertext{Inverting this leading splitting gives the perturbative beat spacing}
\Delta u_{\rm LO}=\frac{D\,r_{\rm RN}^4}{3\mu Q^2b_{\rm RN}}\quad(\mu>0,\ Q\ne0).\label{eq:leadingBeat}
\end{gather}
For $\widehat\mu=0.02$ and $D/M=100$, comparison with the unexpanded rational prescription gives the values in Table~\ref{tab:beat_accuracy}.

\begin{table*}[!t]
\centering
\caption{Leading-order (LO) and unexpanded rational (rat) predictions for the critical-impact-parameter splitting $\Delta b=b_{\rm crit}^{+}-b_{\rm crit}^{-}$ and beat spacing at $\widehat\mu=0.02$ and $D/M=100$. The last column is $100(\Delta u_{\rm rat}/\Delta u_{\rm LO}-1)$.}
\label{tab:beat_accuracy}
\small\setlength{\tabcolsep}{7pt}
\begin{tabular}{crrrrr}
\toprule
$Q/M$ & $(\Delta b/M)_{\rm LO}$ & $(\Delta b/M)_{\rm rat}$ & $\Delta u_{\rm LO}$ & $\Delta u_{\rm rat}$ & shift\\\midrule
0.7 & 0.00291440 & 0.00288797 & 34312.34 & 34626.34 & $0.915\%$\\
1.0 & 0.0150000 & 0.0139721 & 6666.67 & 7157.11 & $7.357\%$\\\bottomrule
\end{tabular}
\end{table*}

Here $\Delta b=b_{\rm crit}^+-b_{\rm crit}^-$ and the last column is $100(\Delta u_{\rm rat}/\Delta u_{\rm LO}-1)$. The rational results reproduce the quoted critical impact parameters, but their displayed numerical precision is conditional on that prescription. The effective metrics control the splitting through first order only; omitted $O(\widehat\mu^2)$ contributions to $\Delta b$ induce $O(\widehat\mu)$ relative corrections in its inverse. The $0.915\%$ and $7.357\%$ shifts are therefore checks of sensitivity to formally higher-order terms, not rigorous error bars for the underlying theory. Figure~8 uses $\Delta u_{\rm rat}$ to normalize its ideal thin-ring baseline. The smaller mode separation than in the double-photon-sphere model of Chen et al. gives a longer modulation scale. In the Maxwell limit the two critical curves merge and the two-frequency beat disappears.

Figure~7 presents the ray-traced image-order intensities, face-on images and separate-mode visibility amplitudes for $Q/M=0.7$. Figure~8 compares the unresolved signals at $Q/M=0.7$ and $1$, including a narrow-ring asymptotic interpretation. In both cases each polarization has a single exterior potential maximum, so summing the modes changes the Fourier signal without creating inter-barrier photon trajectories.

The upper panels of Figure~8 resolve the normalized visibility into direct, lensed and photon contributions. At zero baseline the transform measures accepted flux, so the broad direct component carries most of the signal. As the baseline increases, spatially extended emission loses visibility more rapidly, allowing compact higher-order structures to become relatively important. The locations of minima depend on the angular scales and radial widths of the contributing images. The black total curve is formed by adding the signed contributions before taking the modulus; its minima therefore reflect cancellations as well as diminishing amplitude. A visibility minimum does not imply that the corresponding image has no emission.

The middle panels retain the sign of the numerical photon contribution and multiply it by the square root of baseline. This scaling removes the leading decay of an ideal infinitesimally thin ring, making changes in the oscillation amplitude easier to distinguish from that decay. The ray-traced curves still contain finite-width subrings, unequal image-order weights and source-dependent structure. Their modulation is consequently more complicated than a pure two-ring beat. Comparing the columns tests the charge dependence of both the critical angular radii and the relative brightness of the contributing subrings; their amplitudes alone do not measure the separation of the two polarization light rings.

The lower panels isolate the analytic two-ring interpretation using the flux of the narrowest retained subrings. The fast oscillations are set by the individual critical-curve angular radii, while the slower envelope is set by their difference. The common horizontal variable $u/\Delta u_{\rm rat}$ uses the rational-prescription beat spacings in the comparison above and conceals their difference in absolute baseline. Their perturbative sensitivity is approximately $0.9\%$ and $7.4\%$ relative to the leading-order estimates in these examples. Unequal mode weights prevent exact envelope zeros. These panels illustrate the limiting narrow-ring mechanism and are not a fit to the finite-width numerical curves or a forecast of instrumental detectability. In particular, their extension to three ideal beat periods is separate from the finite-width convergence test on $3000\leq u\leq25000$.

\section{Periodic orbits and gravitational-wave snapshots of EMRIs}

A stellar-mass compact body moving around a much more massive black hole provides an extreme-mass-ratio system. Bound geodesics describe the conservative motion underlying an extreme-mass-ratio inspiral (EMRI); exactly closed periodic orbits are useful reference trajectories, rather than complete inspirals. Radiation reaction gradually changes the orbital energy and angular momentum in a physical EMRI.
Here we determine the periodic geodesics of the ENED background lapse of Eq.~\eqref{eq:lapse}, adopted from Ref.~\cite{dePaula}, together with their approximate quadrupole gravitational-wave snapshots, following the periodic-orbit taxonomy of Ref.~\cite{Levin} and the trajectory-based waveform approach of Ref.~\cite{Babak}. The secondary is neutral and nonspinning, with mass $m_\star\ll M$; its electromagnetic coupling and self-force are neglected.

We use $G=c=1$ and distinguish the NED parameter $\mu$ from $m_\star$. Equation~\eqref{eq:lapse} implies $[\mu]=[M]^2$. With $x=r/M$, $\widehat Q=Q/M$ and $\widehat\mu=\mu/M^2$, the dimensionless form of the background lapse is
\begin{equation}
f(x)=1-\frac{2}{x}+\frac{\widehat Q^2}{x^2}-\frac{2\widehat\mu\widehat Q^4}{5x^6}.\label{eq:existing28}
\end{equation}
The RN limit is $\widehat\mu=0$, while $\widehat Q=0$ gives Schwarzschild for every $\widehat\mu$. The calculations below use Eq.~\eqref{eq:lapse} as the prescribed exterior geometry; its interpretation as the EH-type description additionally requires control of the terms omitted from the effective electromagnetic theory.

For the background spacetime of the adopted model, the equatorial Killing constants are $E=f(r)\dot t$ and $L=r^2\dot\phi$, where the dot denotes proper-time differentiation. The standard timelike normalization and effective-potential construction~\cite{Levin} give
\begin{gather}
\dot r^2=R(r)=E^2-V_{\rm eff}(r),\label{eq:existing29}\\
\shortintertext{For neutral motion, the barrier in this radial equation is}
V_{\rm eff}(r)=f(r)\left(1+\frac{L^2}{r^2}\right).\label{eq:massivePotential}
\end{gather}
These expressions follow from the metric of Eq.~\eqref{eq:geometry}, whose angular coefficient is $r^2$, consistent with the disk equations of Section~VI.
$L$ is specific angular momentum and $E$ is specific energy, with $L>0$ selected without loss of generality.

Circular motion satisfies $R=R'=0$. Its energy and angular momentum are
\begin{gather}
E_c^2=\frac{2f^2}{2f-rf^\prime},\label{eq:existing30}\\
\shortintertext{The same circular-orbit conditions determine the specific angular momentum:}
L_c^2=\frac{r^3f^\prime}{2f-rf^\prime}.\label{eq:circularL}
\end{gather}
Admissibility requires $f>0$, $2f-rf'>0$, and $L_c^2\geq0$ outside the largest positive root $r_h$ of $f=0$. A circular orbit is stable when $V_{\rm eff}''>0$ at fixed $L$, and its ISCO obeys
\begin{equation}
rff''+3ff'-2r(f')^2=0.\label{eq:existing31}
\end{equation}
The outer physical ISCO branch is used. The marginally bound circular orbit instead satisfies $E_c=1$; it is distinct from the ISCO. In particular, Schwarzschild gives $r_{\rm mb}=4M$ and $r_{\rm ISCO}=6M$.

\begin{figure*}[!tp]\centering\includegraphics[width=.94\linewidth,height=.78\textheight,keepaspectratio]{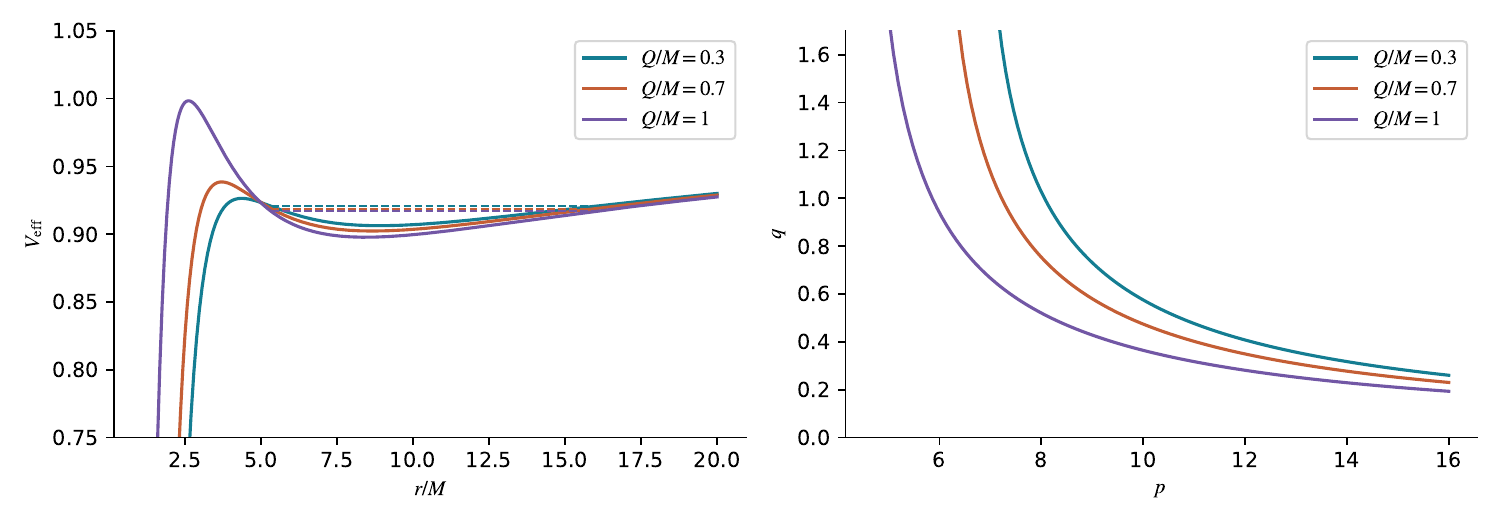}
\caption{Effective potentials and precession for $e=0.5$, $\widehat\mu=0.02$. Left: each potential uses the angular momentum determined at $p=8$ for its own charge; the dashed segment marks its $E^2$ between the turning points $r_p/M=16/3$ and $r_a/M=16$. Right: $q(p)$ on the nonplunging branch for the same eccentricity. Increasing charge lowers $q$ at fixed $p$ for these examples and shifts the separatrix inward.}\end{figure*}

For eccentric bound motion, use the usual semilatus rectum $p$ and eccentricity $0<e<1$ to parametrize the turning points~\cite{Fujita}:
\begin{gather}
r_p=\frac{pM}{1+e},\label{eq:existing32}\\
\shortintertext{The apocenter of the same eccentric orbit is}
r_a=\frac{pM}{1-e},\label{eq:apocenter}\\
\shortintertext{Both turning points are combined in the standard radial-anomaly parametrization~\cite{Fujita}:}
r(\chi)=\frac{pM}{1+e\cos\chi}.\label{eq:anomalyRadius}
\end{gather}
Writing $f_p=f(r_p)$ and $f_a=f(r_a)$, the two turning-point conditions determine
\begin{gather}
L^2=\frac{f_a-f_p}{f_p/r_p^2-f_a/r_a^2},\label{eq:existing33}\\
\shortintertext{Substitution at the pericenter then fixes the specific energy:}
E^2=f_p\left(1+\frac{L^2}{r_p^2}\right).\label{eq:boundEnergy}
\end{gather}
We retain the exterior bound branch with $E<1$, $r_p>r_h$ and $R>0$ between the turning points. Its separatrix is found from $R(r_p)=R'(r_p)=0$, where the inner turning point merges with an unstable circular orbit. Thus $p>p_{\rm sep}(e)$ is required on this branch, rather than simply $r_p>r_{\rm ISCO}$. Eccentric orbits can reach inside the ISCO without plunging.

Using the standard period-and-frequency description of bound geodesics~\cite{Schmidt,Fujita}, the azimuthal advance is
\begin{gather}
\Delta\phi=2\int_{r_p}^{r_a}\frac{L\,dr}{r^2\sqrt{R(r)}},\label{eq:existing34}\\
\shortintertext{The corresponding radial period in asymptotically normalized coordinate time is}
T_r=2\int_{r_p}^{r_a}\frac{E\,dr}{f(r)\sqrt{R(r)}}.\label{eq:radialPeriod}
\end{gather}
The substitution in Eq.~\eqref{eq:anomalyRadius} removes the integrable turning-point singularities. The fundamental frequencies and rotation number are
\begin{gather}
\Omega_r=\frac{2\pi}{T_r},\label{eq:existing35}\\
\shortintertext{The azimuthal frequency measures the angle accumulated during that period:}
\Omega_\phi=\frac{\Delta\phi}{T_r},\label{eq:azimuthalFrequency}\\
\shortintertext{Their ratio gives the precession parameter used in the periodic-orbit taxonomy~\cite{Levin}:}
q=\frac{\Omega_\phi}{\Omega_r}-1=\frac{\Delta\phi}{2\pi}-1.\label{eq:rotationNumber}
\end{gather}
Following Ref.~\cite{Levin}, a closed orbit has rational $q=w+v/z$, where $z$ counts radial cycles before closure, $w$ counts additional complete azimuthal revolutions per radial cycle, and $v$ specifies the order in which the leaves are visited. For $z>1$, $0<v<z$ and $v,z$ are coprime. The one-leaf orbit has $z=1,v=0$. Closure therefore requires
\begin{equation}
z\Delta\phi=2\pi\,[z(1+w)+v].\label{eq:existing36}
\end{equation}
The families used below are $(z,w,v)=(2,0,1),\allowbreak(3,0,2),\allowbreak(1,1,0)$, corresponding to $q=1/2,2/3,1$. Their orbital periods are $zT_r$. As the separatrix is approached from above, the time spent near the unstable circular orbit increases, producing zoom-whirl motion and an increasing $q$.

\begin{figure*}[!tp]\centering\includegraphics[width=.94\linewidth,height=.78\textheight,keepaspectratio]{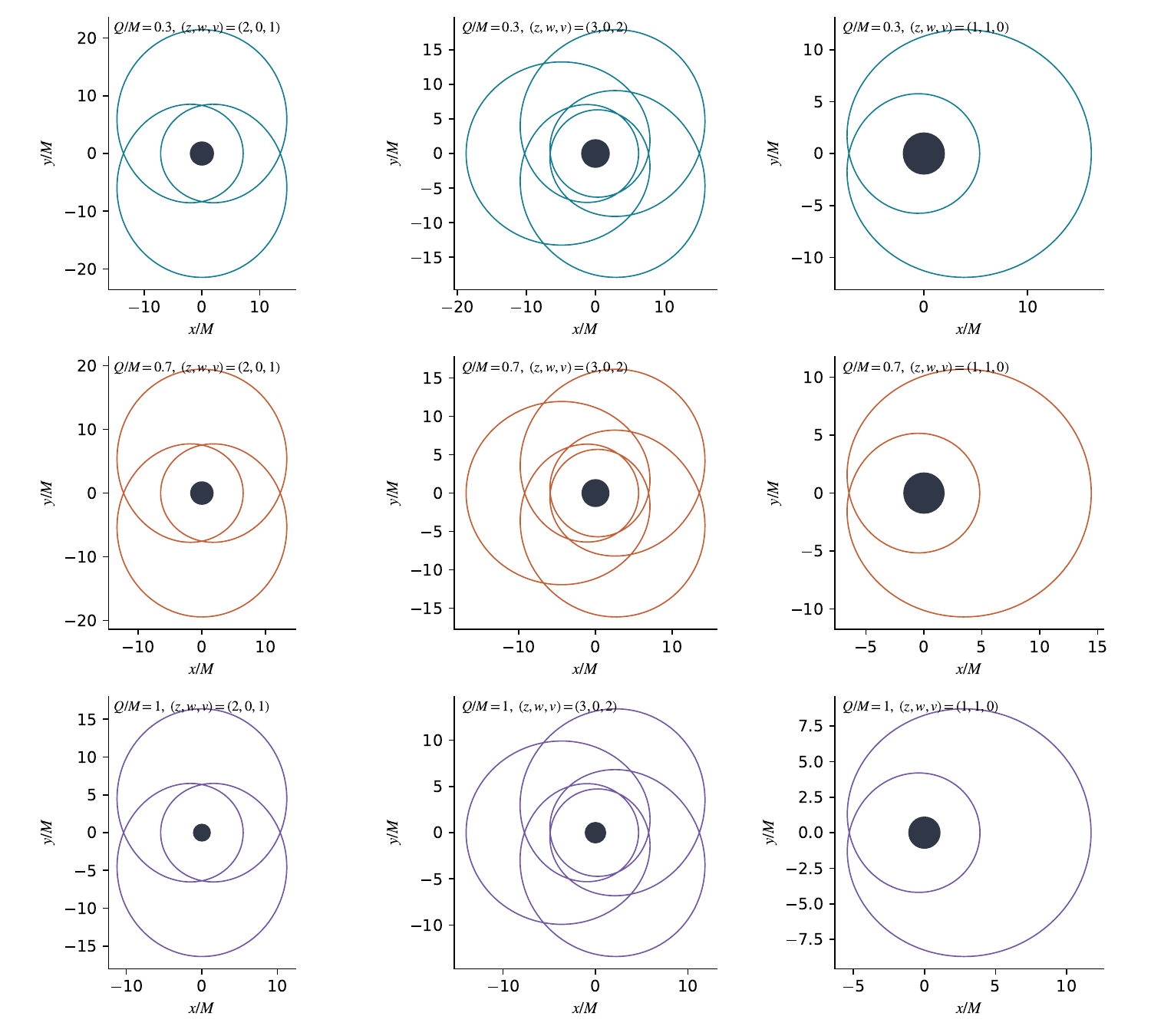}
\caption{Nine closed equatorial timelike geodesics of Eq.~\eqref{eq:lapse}, with $e=0.5$ and $\widehat\mu=0.02$. Rows show $\widehat Q=0.3,0.7,1.0$, and columns show $(z,w,v)=(2,0,1),(3,0,2),(1,1,0)$. Each curve spans exactly $z$ radial cycles, using the unrounded values underlying Table~\ref{tab:closed_orbits}. The filled central disk denotes the horizon. These are coordinate-plane projections $x=r\cos\phi$, $y=r\sin\phi$; they are not embeddings of proper spatial distance.}\end{figure*}

\begin{table}[ht]\centering
\caption{Exterior boundaries calculated from Eq.~\eqref{eq:lapse}, for $\widehat\mu=0.02$. The eccentric separatrix uses $e=0.5$; $p$ is dimensionless.}
\label{tab:exterior_boundaries}
\begin{tabular}{cccc}\toprule
$\widehat Q$ & $r_h/M$ & $r_{\rm ISCO}/M$ & $p_{\rm sep}(0.5)$\\\midrule
0.3 & 1.953942 & 5.862784 & 6.839852\\
0.7 & 1.714299 & 5.185265 & 6.046771\\
1.0 & 1.077097 & 4.000638 & 4.638533\\\bottomrule
\end{tabular}\end{table}
The $\widehat Q=1$ example is nonextremal at positive $\widehat\mu$. Its horizon value in Table~\ref{tab:exterior_boundaries} follows the square-root displacement derived in Section~II, so a regular first-order Taylor shift around the extremal RN horizon is not applicable. The outer photon and ISCO radii remain nondegenerate exterior scales with regular small-coupling expansions.

Figure~9 connects the effective-potential barrier to the precession of eccentric motion. In the left panel the same turning-point radii are imposed across the charge comparison; $E$ and $L$ must consequently change with $Q$. In the right panel, decreasing $p$ enhances the strong-field part of the orbit and raises $q$. All plotted trajectories lie above their respective eccentric separatrices.

\begin{table}[ht]\centering\small
\caption{Closed-orbit examples for $e=0.5$, $\widehat\mu=0.02$. The values of $p$ solve the indicated rational-$q$ condition; $E$ and $L$ follow from Eqs.~\eqref{eq:existing33}-\eqref{eq:boundEnergy}. Rounded values are shown.}
\label{tab:closed_orbits}
\begin{tabular}{ccrrrr}\toprule
$\widehat Q$ & $(z,w,v)$ & $p$ & $E$ & $L/M$ & $T_r/M$\\\midrule
0.3 &(2,0,1)&10.725743&0.967964&3.900874&461.247\\
0.3 &(3,0,2)&9.351931&0.964071&3.760220&405.779\\
0.3 &(1,1,0)&8.063331&0.959727&3.644174&370.036\\
0.7 &(2,0,1)&9.716672&0.964744&3.687638&404.806\\
0.7 &(3,0,2)&8.431557&0.960291&3.545453&354.784\\
0.7 &(1,1,0)&7.222194&0.955245&3.426242&322.054\\
1.0 &(2,0,1)&8.190872&0.958397&3.350722&323.745\\
1.0 &(3,0,2)&6.999125&0.952482&3.197723&279.547\\
1.0 &(1,1,0)&5.859115&0.945381&3.062392&248.626\\\bottomrule
\end{tabular}\end{table}
Table~\ref{tab:closed_orbits} and Fig.~10 compare the same closure class across different charges. Since $q$ is fixed, $p$ is adjusted independently rather than imposing common $E$ or $L$. Increasing $Q$ shifts these particular closed orbits inward and reduces their radial periods. The result is a comparison of families in different geometries, not the evolution of one orbit as charge changes.

The two-leaf and three-leaf trajectories close after two and three radial cycles, respectively. The one-leaf whirl orbit completes two azimuthal revolutions during a single radial oscillation. Each curve is calculated from $d\phi/d\chi=L|dr/d\chi|/(r^2\sqrt R)$ and $dt/d\chi=E|dr/d\chi|/(f\sqrt R)$; no orbit is drawn by imposing an artificial azimuthal frequency. Charge changes the radius and timing needed to realize a prescribed rational rotation number.

Following the numerical-kludge construction~\cite{Babak}, we map the trajectory to auxiliary Cartesian coordinates $\mathbf{x}=(r\cos\phi,r\sin\phi,0)$ and form its trace-free mass quadrupole:
\begin{gather}
I_{ij}=m_\star\left(x_ix_j-\frac13\delta_{ij}r^2\right),\label{eq:existing37}\\
\shortintertext{In the slow-motion quadrupole prescription~\cite{Babak}, the transverse-traceless strain is}
h_{ij}^{\rm TT}(t)=\frac{2G}{c^4D}\ddot I_{ij}^{\rm TT}(t-D/c),\label{eq:quadrupoleWave}
\end{gather}
where $D$ is the source distance and derivatives use coordinate time. This combines a relativistic geodesic with a flat-space radiation formula: it is a quadrupole kludge, not a solution of the coupled Einstein-NED perturbation equations. The background NED field affects the trajectory through $f(r)$, while its radiative perturbations and radiation reaction are omitted. Ref.~\cite{Babak} motivates this construction, but Kerr waveform accuracy estimates from that work do not establish accuracy for the present NED geometry.

\begin{figure*}[!tp]\centering\includegraphics[width=.94\linewidth,height=.78\textheight,keepaspectratio]{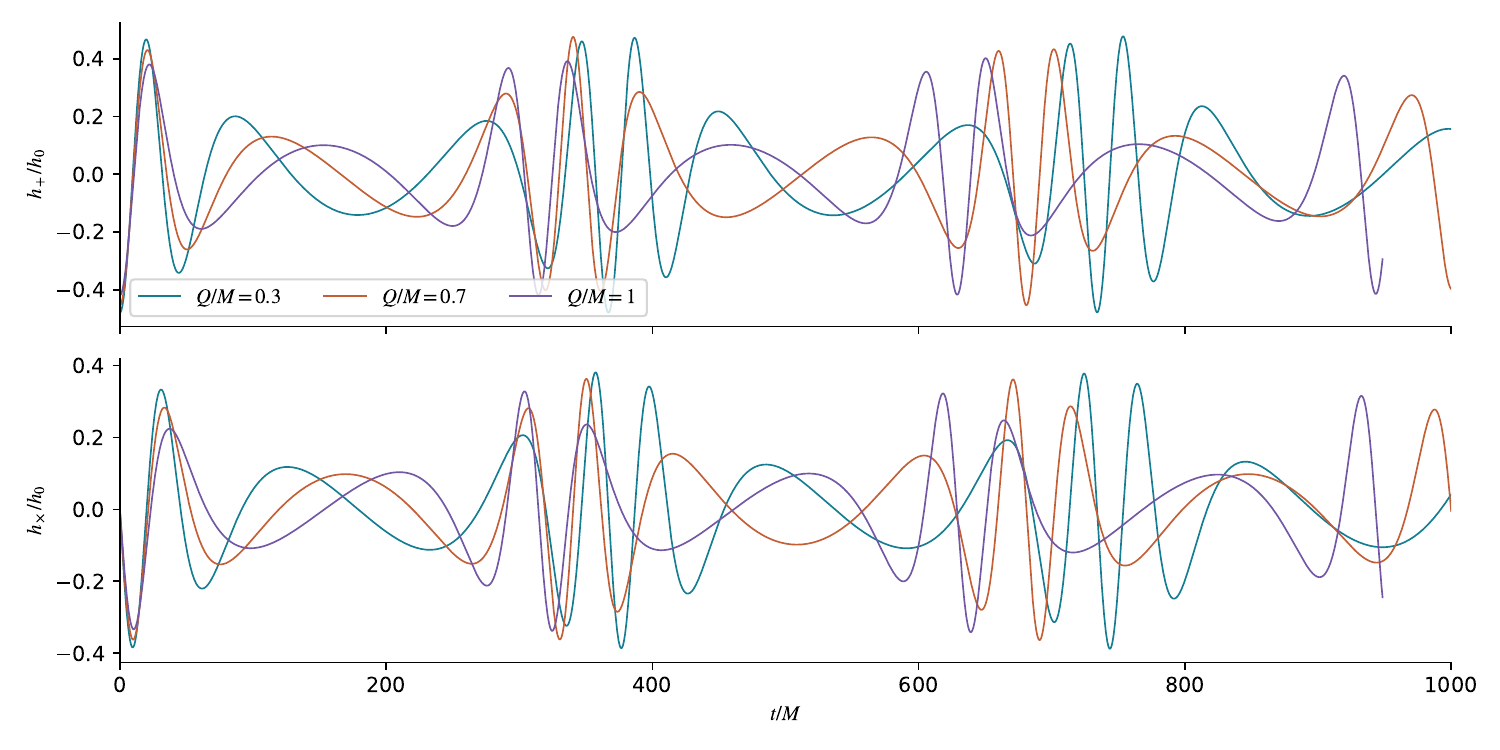}
\caption{Quadrupole waveform snapshots for $p=8$, $e=0.5$, $\widehat\mu=0.02$, and $\iota=\pi/3$. Charge varies while the turning-point radii remain fixed. Each orbit starts at pericenter with $t=\phi=0$, and $E,L$ are recomputed for its charge. The common window is $0\leq t/M\leq1000$. These generally nonclosed bound trajectories show charge-dependent timing and phase; $h_0=Gm_\star/(c^2D)$.}\end{figure*}

For an observer at inclination $\iota$ with $\mathbf n=(0,\sin\iota,\cos\iota)$, choose polarization vectors $\mathbf e_1=(1,0,0)$ and $\mathbf e_2=(0,\cos\iota,-\sin\iota)$. Then
\begin{gather}
h_+=\frac{Gm_\star}{c^4D}\frac{d^2}{dt^2}\left(x^2-y^2\cos^2\iota\right),\label{eq:existing38}\\
\shortintertext{The orthogonal cross polarization follows from the same projection:}
h_\times=\frac{2Gm_\star\cos\iota}{c^4D}\frac{d^2}{dt^2}(xy).\label{eq:waveCross}
\end{gather}
In geometrized coordinates we normalize both polarizations by $h_0=Gm_\star/(c^2D)$. Coordinate velocities are $dr/dt=(f/E)\dot r$ and $d\phi/dt=fL/(Er^2)$. Analytic coordinate accelerations avoid finite-difference artifacts at pericenter:
\begin{gather}
\frac{d^2r}{dt^2}=\frac{ff^\prime}{E^2}R+\frac{f^2}{2E^2}R^\prime,\label{eq:existing39}\\
\shortintertext{The coordinate-time angular equation gives the corresponding azimuthal acceleration:}
\frac{d^2\phi}{dt^2}=\frac{d\phi}{dt}\left(\frac{f^\prime}f-\frac2r\right)\frac{dr}{dt}.\label{eq:angularAcceleration}
\end{gather}
In this equation $R'$ is evaluated at fixed $E,L$. Equations~\eqref{eq:existing38}-\eqref{eq:waveCross} are then evaluated using $d^2(x_ix_j)/dt^2=2v_iv_j+x_i a_j+a_i x_j$.

Figure~11 demonstrates waveform phase and amplitude changes at fixed geometric orbit size. It uses a different comparison from Fig.~10: $p$ and $e$ are held fixed, so the rotation numbers are generally irrational. Their $T_r/M$ values are $369.114$, $340.757$, and $316.156$ for $\widehat Q=0.3,0.7,1.0$, respectively. Repeated pericenter passages produce enhanced emission, and the changing azimuth of successive passages changes the polarization pattern. At fixed constants of motion, the conservative waveform has harmonics
\begin{equation}
\omega_{mn}=m\Omega_\phi+n\Omega_r,\qquad m,n\in\mathbb Z.\label{eq:existing40}
\end{equation}
The integers label Fourier components, not positive counts of orbital cycles. Negative $n$ gives radial sidebands below the corresponding azimuthal harmonic. A physical positive frequency uses $\omega_{mn}>0$, with conjugate components ensuring a real strain. These orbital frequencies are distinct from black hole quasinormal-mode frequencies.

\begin{figure*}[!tp]\centering\includegraphics[width=.94\linewidth,height=.78\textheight,keepaspectratio]{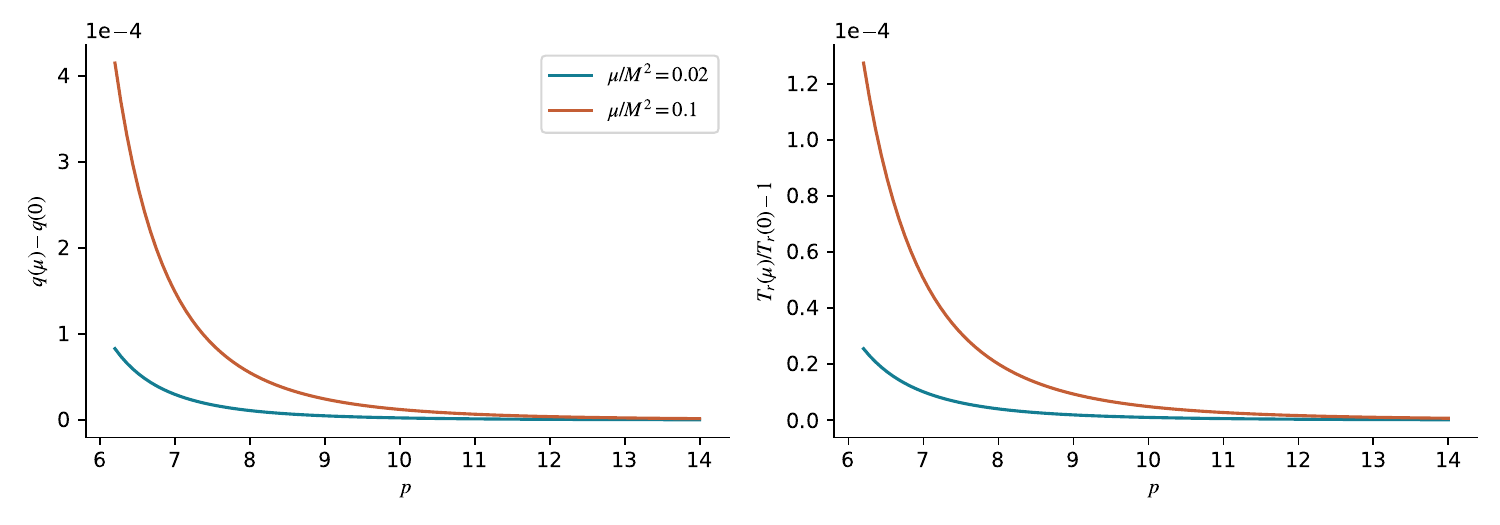}
\caption{Isolating the NED correction at fixed $\widehat Q=1$, $e=0.5$. Left: $q(\widehat\mu)-q(0)$. Right: $T_r(\widehat\mu)/T_r(0)-1$. The reference $\widehat\mu=0$ is RN at identical $p,e,Q$, and both couplings lie on the nonplunging branch over the plotted interval. Difference plots resolve the small additional NED effect without conflating it with the much larger charge effect.}\end{figure*}

For scale, $M=10^6M_\odot$ gives $GM/c^3\simeq4.9255\,$s, and the $\widehat Q=0.7$, $p=8$ example has $T_r\simeq1678\,$s. With $m_\star=10M_\odot$ and $D=1\,$Gpc, $m_\star/M=10^{-5}$ and $h_0\simeq4.79\times10^{-22}$. These are illustrative scales, neglecting cosmological redshift, rather than a detector sensitivity forecast.

{\looseness=1
The $r^{-6}$ correction is most important near the inner part of an orbit. Figure~12 shows that its influence rapidly weakens as $p$ grows, for the examples considered. Positive coupling increases $q$ and $T_r$ relative to RN over this plotted interval. The difference is small for $\widehat\mu=0.02$; the visually substantial changes in Figs.\,9-11 are principally charge effects. No parameter constraint or universal monotonicity claim follows from this limited sample.\par}

The trajectories are obtained by solving the turning-point and rational-$q$ conditions, with the unrounded values used for the figures. As consistency checks, $Q=0$ recovers $r_h/M=2$, $r_{\rm ISCO}/M=6$, and $p_{\rm sep}=6+2e$, while $\mu=0$ recovers RN. The computed examples satisfy the bound exterior conditions and close in azimuth within $3\times10^{-11}$ radians before rounding.

The admissible energy-angular-momentum domain can also be displayed without fixing an eccentricity. For each $L$, let $r_s(L)$ and $r_u(L)$ denote the stable and unstable circular radii on the exterior branch, and define $E_s(L)=\sqrt{V_{\rm eff}(r_s)}$ and $E_u(L)=\sqrt{V_{\rm eff}(r_u)}$. The nonplunging eccentric bound domain, with the prescribed final energy $E_{\max}=0.999999$, obeys
$L_{\rm ISCO}<L<L_{\max}$.
At fixed $L$, the stable circular orbit sets the lower energy boundary:
$E_s(L)<E\leq0.999999$. An exterior nonplunging orbit must also remain below the unstable circular barrier:
$E<E_u(L)$.

The stable circular boundary $E=E_s$ is allowed if circular orbits are included, but is excluded for nonzero eccentricity. The unstable boundary $E=E_u$ is the separatrix and is excluded from finite-period eccentric motion. The ISCO is the cusp where the two circular boundaries meet.

\begin{figure}[!htbp]\centering\includegraphics[width=.98\linewidth,height=.78\textheight,keepaspectratio]{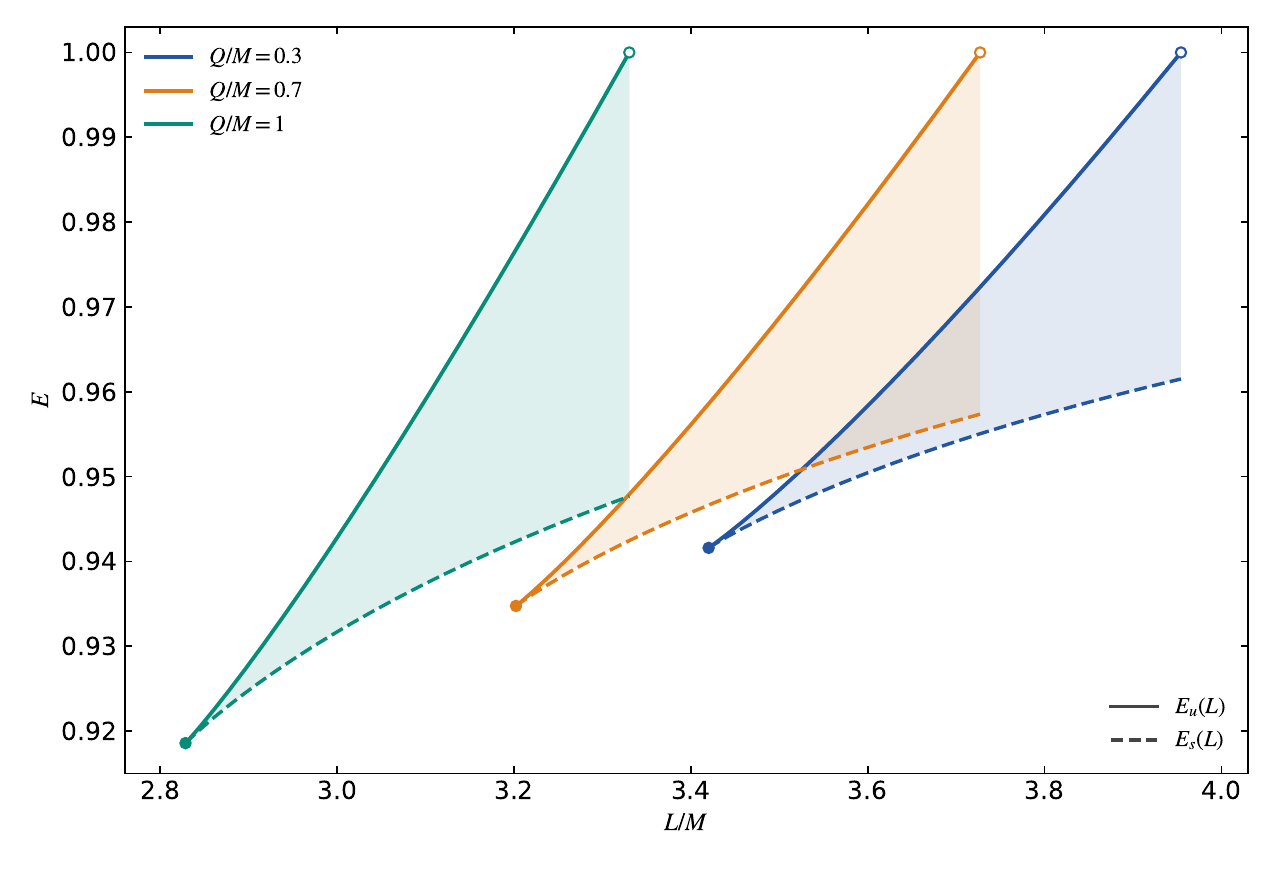}
\caption{Restricted strong-field bound-orbit regions for the ENED background, at $\widehat\mu=0.02$. Blue, orange and green correspond to $Q/M=0.3,0.7,1.0$, respectively. Each shaded region spans $L_{\rm ISCO}<L<L_{\rm crit}$ between the stable circular energy $E_s(L)$ (dashed) and the unstable circular energy $E_u(L)$ (solid). Filled dots mark the ISCOs. Open circles mark the cutoff endpoints $E_u(L_{\rm crit})=0.999999$, just below the marginally bound energy; they are not exactly marginally bound orbits. Finite-period eccentric motion excludes both circular boundaries. The permitted domain with $L>L_{\rm crit}$ remains present but is omitted from this restricted view. No eccentricity is fixed.}\end{figure}

Figure~13 overlays the three charge-dependent strong-field domains in one linear-axis panel. Dashed lower boundaries are stable circular orbits, solid upper boundaries are unstable circular separatrices, and each filled ISCO dot is their common cusp. Increasing charge shifts the displayed cusps and upper endpoints to smaller angular momenta and lowers the ISCO energy in these examples. The open upper endpoints enforce $E_{\max}=0.999999$ rather than unity, so the plotted right edges are $L_{\rm crit}$, just below $L_{\rm mb}$. They are plotting limits for this restricted comparison, not upper limits on all bound motion. The permitted energy interval depends on angular momentum. When $L_{\rm ISCO}<L<L_{\rm crit}$, the unstable circular orbit sets the upper boundary; increasing energy past that boundary removes the barrier protecting the exterior potential well from plunge. Here $L_{\rm crit}$ satisfies $E_u(L_{\rm crit})=E_{\max}$. When $L>L_{\rm crit}$, the imposed energy cutoff is the active upper boundary instead; this continuation is omitted from Figure~13, while its full-domain endpoint is retained in Table~\ref{tab:limits_domain} and the following analysis. Below $E_s(L)$, no motion in the exterior bound well is possible.

\begin{table*}[t]
	\centering
	\small
	\caption{Limits of the admissible domain for $\widehat\mu=0.02$ and $E_{\max}=0.999999$. $L_{\rm crit}$ is the switch between the separatrix and energy-cutoff upper boundaries; $L_{\rm mb}$ belongs to the unstable marginally bound circular orbit with $E_u=1$.}
	\label{tab:limits_domain}
	\begin{tabular}{crrrrr}
		\toprule
		$\widehat Q$ & $E_{\rm ISCO}$ & $L_{\rm ISCO}/M$ & $L_{\rm crit}/M$ & $L_{\rm mb}/M$ & $L_{\max}/M$ \\
		\midrule
		0.3 & 0.941586629 & 3.420061431 & 3.954206169 & 3.954213986 & 707.108309 \\
		0.7 & 0.934744345 & 3.202224379 & 3.727032286 & 3.727039211 & 707.108026 \\
		1.0 & 0.918565029 & 2.828500746 & 3.330360311 & 3.330365701 & 707.107665 \\
		\bottomrule
	\end{tabular}
\end{table*}
The upper angular-momentum endpoint is fixed by the stable circular energy reaching the chosen cutoff: $E_s(L_{\max})=0.999999.$
At this endpoint the eccentric domain collapses to a circular orbit. Its radius is approximately $r_s/M=499999.25$ in all three examples. This large radius follows from choosing an energy extremely close to the rest-mass threshold. In the weak-field limit, $1-E_s\simeq M/(2r_s)$ and $L_s/M\simeq\sqrt{r_s/M}$, explaining why $L_{\max}/M$ is close to $707.1$.

The commonly used interval $L_{\rm ISCO}<L<L_{\rm mb}$ selects the strong-field families for which the unstable circular barrier has energy below unity. It does not describe the entire bound-orbit domain: nonplunging, weakly bound orbits also exist for $L>L_{\rm mb}$. Figure~13 deliberately displays only the near-ISCO, strong-field subset; the full-domain values $L_{\max}$ in Table~\ref{tab:limits_domain} describe the omitted weakly bound continuation. Its cutoff endpoints $L_{\rm crit}$ differ slightly from the marginally bound values $L_{\rm mb}$ listed in that table.

An interior point of the shaded region generically corresponds to a precessing, nonclosed bound orbit. Exactly periodic orbits are the rational-$q$ subset of that region, selected by Eq.~\eqref{eq:existing36}; the nine examples in Table~\ref{tab:closed_orbits} lie in the allowed domain. The shading therefore indicates admissible bound motion and must not be interpreted as saying that every plotted $(E,L)$ combination closes after a finite number of radial cycles.

\section{Accretion disk dynamics and thermal radiation}

We next adopt a stationary, geometrically thin disk whose neutral matter moves on nearly circular geodesics. Stationarity refers to time-independent disk profiles; the gas has nonzero azimuthal velocity, with $u_{\rm circ}^\alpha=u^t(1,0,0,\Omega)$ in the circular-orbit approximation. This orbiting, optically thick thermal source is a separate configuration from the static, optically thin emitters in Sections~III-IV. Their different velocity fields, emissivities, opacities, and inner-boundary prescriptions preclude interpreting their brightness and spectral results as predictions for one fixed accretion flow. Its self-gravity, radial pressure support and advective heat transport are neglected, and the inner edge is the background ISCO. The disk is radiatively efficient, with equal local fluxes from its two faces and zero torque at the inner edge. This is the standard conservation-law thin-disk setting of Page and Thorne~\cite{PageThorne}; the formulas below specialize it to the ENED background lapse~\cite{dePaula}, given in Eq.~\eqref{eq:lapse}. A conducting, magnetically coupled plasma would need additional force and stress equations and is not represented by this neutral-matter idealization.

Using the neutral-geodesic variational principle~\cite{Levin}, the equatorial Lagrangian for the adopted background is
\begin{gather}
\mathcal L_{\rm geo}=\frac12\left[-f(r)\dot t^2+\frac{\dot r^2}{f(r)}+r^2\dot\phi^2\right],\label{eq:accLag}\\
\shortintertext{Stationarity of the geodesic action gives the Euler-Lagrange equations:}
\frac{d}{d\tau}\frac{\partial\mathcal L_{\rm geo}}{\partial\dot x^\alpha}-\frac{\partial\mathcal L_{\rm geo}}{\partial x^\alpha}=0.\label{eq:eulerLagrange}
\end{gather}
The coordinate $\phi$ is cyclic, as is $t$, giving $E=f\dot t$ and $L=r^2\dot\phi$. The angular metric coefficient in Eq.~\eqref{eq:geometry} is $r^2$; no spin parameter enters this spherical geometry. The normalization condition recovers the timelike potential already introduced in Eq.~\eqref{eq:existing29}; it is not a new independent radial equation.

For circular motion, the radial Euler-Lagrange equation reduces to $f'\dot t^2=2r\dot\phi^2$. Defining $\Omega=\dot\phi/\dot t$ and using the timelike normalization gives
\begin{gather}
\Omega^2=\frac{f^\prime}{2r}=\frac{M}{r^3}-\frac{Q^2}{r^4}+\frac{6\mu Q^4}{5r^8},\label{eq:accOmega}\\
\shortintertext{The circular material four-velocity is normalized using the background spacetime metric:}
u^t=\frac{1}{\sqrt{H(r)}},\label{eq:circularClock}\\
\shortintertext{It is convenient to collect these normalization factors in the function}
H(r)\equiv f-\frac{rf^\prime}{2}.\label{eq:circularH}
\end{gather}
For the lapse in Eq.~\eqref{eq:lapse}, this denominator and the circular constants become
\begin{gather}
H(r)=1-\frac{3M}{r}+\frac{2Q^2}{r^2}-\frac{8\mu Q^4}{5r^6},\label{eq:accEL}\\
\shortintertext{The specific energy is $E(r)=f(r)/\sqrt{H(r)}$, while the corresponding specific angular momentum is}
L(r)=\frac{\sqrt{Mr-Q^2+6\mu Q^4/(5r^4)}}{\sqrt{H(r)}},\label{eq:diskAngularMomentum}\\
\shortintertext{The combination entering the disk conservation laws~\cite{PageThorne} can then be written as}
\mathcal Z(r)\equiv E-\Omega L=\sqrt{H(r)},\label{eq:diskZ}\\
\shortintertext{Differentiating the circular-orbit relations gives the useful identity}
E^\prime=\Omega L^\prime.\label{eq:diskDerivativeIdentity}
\end{gather}
The circular energy and angular momentum specialize Eqs.~\eqref{eq:existing30}-\eqref{eq:circularL}. The last identity follows by differentiating the circular solutions and is essential to simplifying the disk conservation equations. Physical circular disk motion requires $f>0$, $H>0$, nonnegative $L^2$, and radial stability. Figure~14 displays $\Omega,E,L$ only outside each model's own ISCO, together with the corresponding RN-subtracted fractional corrections at the same radius.

\begin{figure*}[!tp]\centering\includegraphics[width=.94\linewidth,height=.78\textheight,keepaspectratio]{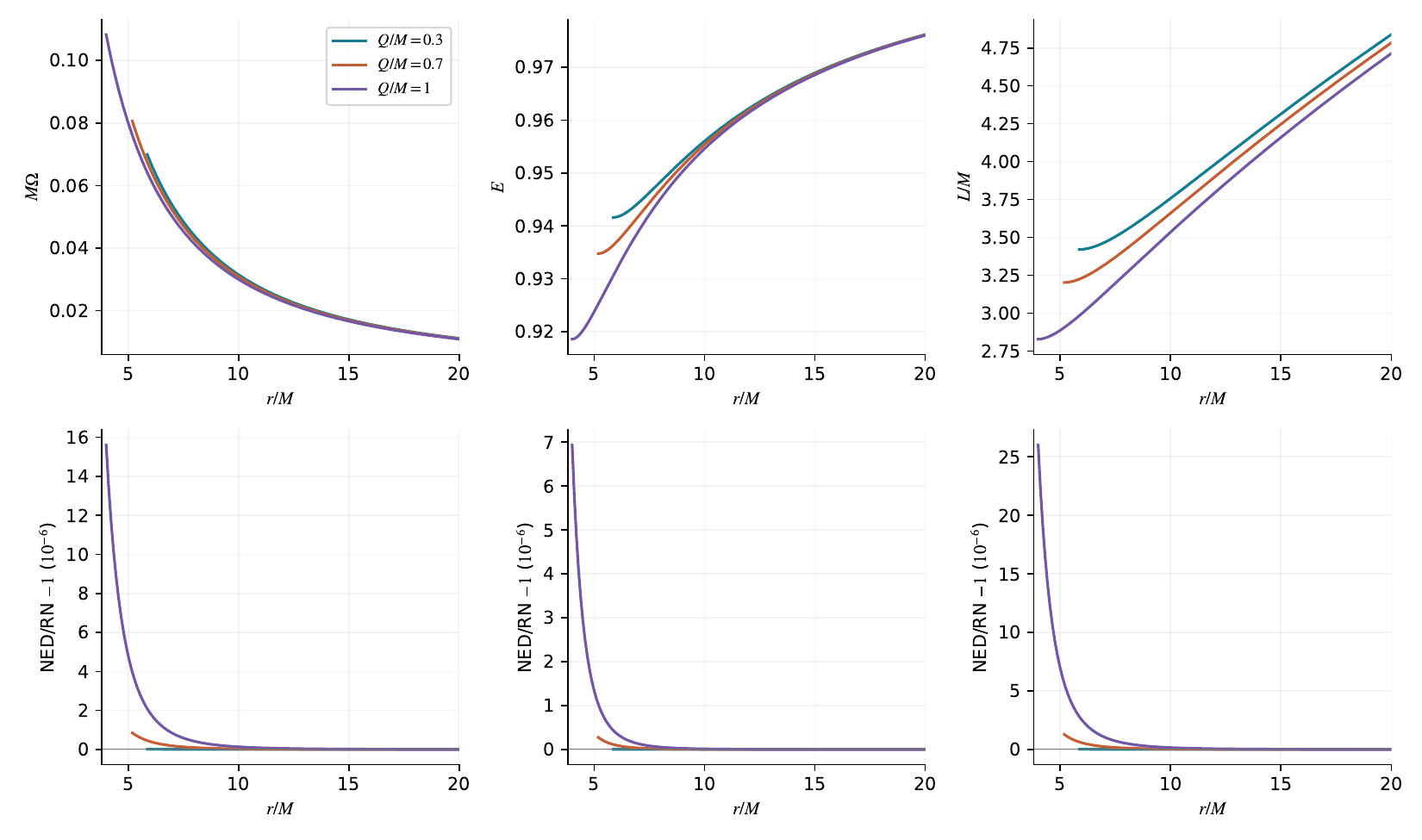}\caption{Circular disk kinematics at $\widehat\mu=0.02$. Top row: $M\Omega$, $E$ and $L/M$ for $Q/M=0.3,0.7,1.0$, each plotted only at and outside its own ISCO. Bottom row: fractional NED/RN changes at the same radius and charge, scaled by $10^{-6}$. These panels separate the much smaller NED contribution from the charge dependence.}\end{figure*}

The ISCO follows from the marginal-stability condition in Eq.~\eqref{eq:existing31}. An equivalent form is $dL^2/dr=0$ on the physical circular branch. Writing $x=r/M$, $q=Q/M$, and $v=\mu/M^2$, the explicit polynomial condition is
\begin{equation}
\begin{split}
25x^{11}-150x^{10}+225q^2x^9-100q^4x^8
-120vq^4x^6-10vq^4x^5\\
+120vq^6x^4-96v^2q^8=0.
\end{split}\label{eq:accISCO}
\end{equation}
To obtain it, write $L^2/M^2=N/H$ with $N=x-q^2+6vq^4/(5x^4)$ and impose $N'H-NH'=0$. Multiplication by $25x^{11}$ yields Eq.~\eqref{eq:accISCO}; derivatives in this sentence are with respect to $x$. The term quadratic in $v$ arises from algebraic products of the prescribed lapse and its derivatives, not from adding a higher-order NED term to the Lagrangian. One selects the exterior root connected to the ordinary outer stable branch. Schwarzschild gives $x_{\rm ISCO}=6$; for extremal RN the admissible outer root is $x_{\rm ISCO}=4$.

For a zero-torque inner boundary, the binding-energy conversion efficiency~\cite{PageThorne} is
\begin{equation}
\eta=1-E(r_{\rm ISCO}).\label{eq:accEta}
\end{equation}
This is an emitted-energy efficiency for a zero-torque inner edge, before deducting photon capture or accounting for returning radiation. Equations~\eqref{eq:accOmega}-\eqref{eq:accISCO} keep the dependence on mass, charge and coupling explicit. Figure~15 shows both the absolute ISCO and efficiency maps and their differences from RN at the same charge. The difference panels reveal the small NED contribution that is obscured by the charge dependence in the absolute maps. The upper panels connect the displacement of the disk inner edge to the emitted binding-energy efficiency over the parameter domain. The lower panels subtract RN at the same charge and therefore isolate the nonlinear coupling response; they complement the single-parameter radial profiles rather than repeating them. A smaller ISCO and a larger emitted efficiency characterize a more compact, more strongly bound emitting disk, within the zero-torque assumptions.

\begin{figure*}[!tp]\centering\includegraphics[width=.94\linewidth,height=.78\textheight,keepaspectratio]{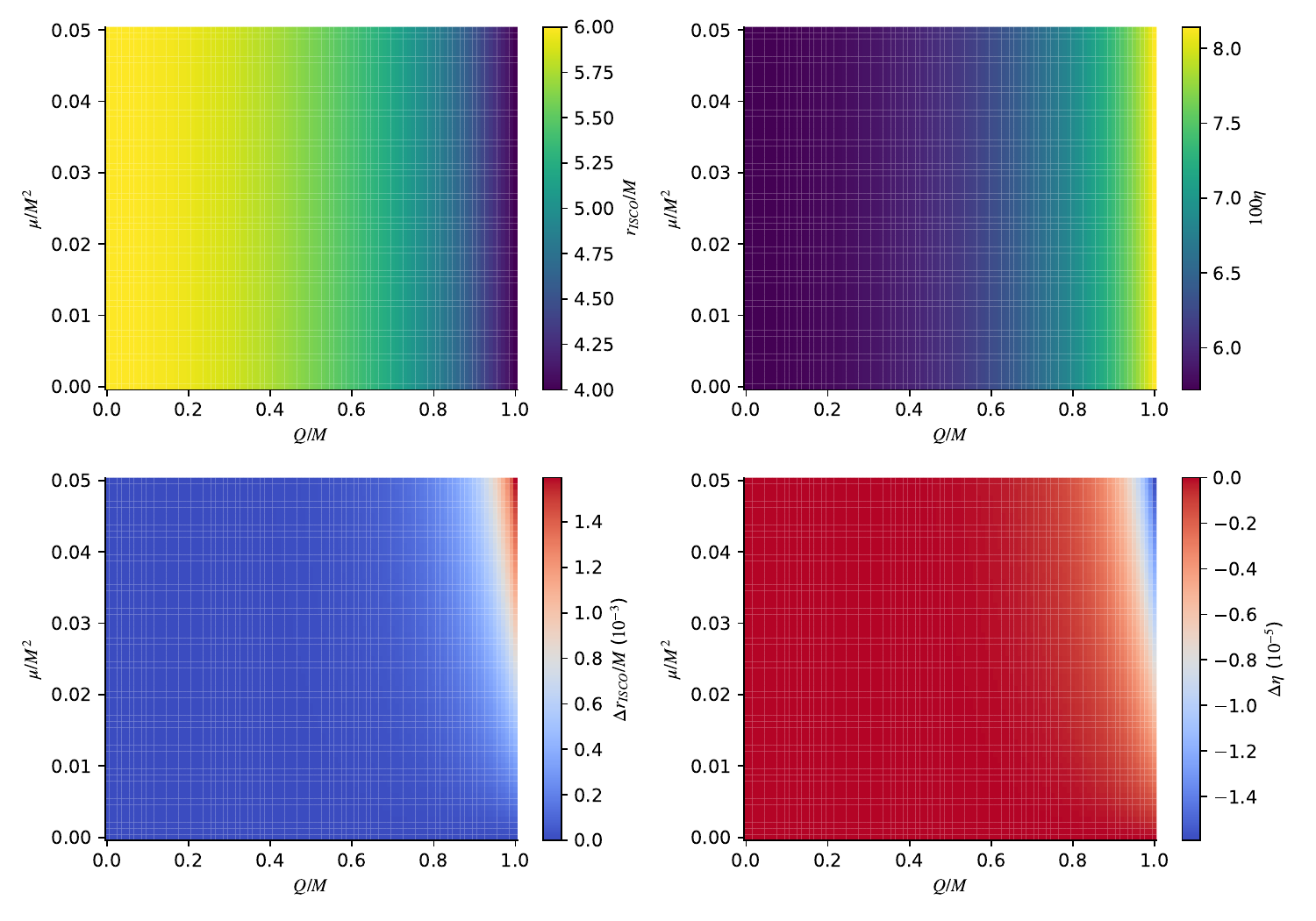}\caption{ISCO and efficiency across $0\leq Q/M\leq1$, $0\leq\mu/M^2\leq0.05$. Upper panels show the absolute $r_{ISCO}/M$ and efficiency in percent. Lower panels isolate their changes relative to RN at identical charge: $\Delta r_{ISCO}/M$ in units of $10^{-3}$ and $\Delta\eta$ in units of $10^{-5}$. The displayed outer root belongs to the background neutral timelike branch. The residual maps expose the NED response without attributing the charge effect to the coupling.}\end{figure*}

Let $\Sigma$ be the vertically integrated rest-mass surface density and $u^r<0$ the small inward mean velocity. Mass conservation gives $\dot M=-2\pi r\Sigma u^r>0$, independent of radius. In local proper-height coordinates $z$ near the equatorial plane, the relevant volume factor is $\sqrt{-g_{(t,r,z,\phi)}}=r$. It is not the spherical-coordinate four-dimensional factor $r^2\sin\theta$. Define the outward transport torque $\mathcal T=2\pi rW^r{}_{\phi}$, where $W^r{}_{\phi}$ is the vertically integrated stress. The height-integrated energy and angular-momentum conservation laws of Page and Thorne~\cite{PageThorne} give
\begin{gather}
\frac{d}{dr}\left(\dot M E-\Omega\mathcal T\right)=4\pi rFE,\label{eq:accConserve}\\
\shortintertext{The corresponding angular-momentum balance is}
\frac{d}{dr}\left(\dot M L-\mathcal T\right)=4\pi rFL.\label{eq:angularConservation}
\end{gather}
Here $F(r)$ is the emitted bolometric energy flux from one face, measured in the orbiting gas frame. The factor $4\pi$ accounts for both faces. Subtracting $\Omega$ times the angular-momentum equation from the energy equation, and using $E'=\Omega L'$, gives $-\Omega'\mathcal T=4\pi rF\mathcal Z$. Combining this with $\mathcal Z'=-\Omega'L$ produces $(\mathcal T\mathcal Z)'=\dot M\mathcal Z L'$. The zero-torque condition $\mathcal T(r_{\rm ISCO})=0$ then fixes the integration constant.

The torque and the local flux therefore are
\begin{gather}
\mathcal T(r)=\frac{\dot M}{\mathcal Z(r)}\int_{r_{\rm ISCO}}^r\mathcal Z(s)L^\prime(s)\,ds,\label{eq:diskTorque}\\
\shortintertext{The shear converts angular-momentum transport into the local one-face emitted flux~\cite{PageThorne}:}
F(r)=-\frac{\dot M\Omega^\prime(r)}{4\pi r\mathcal Z(r)^2}\int_{r_{\rm ISCO}}^r\mathcal Z(s)L^\prime(s)\,ds.\label{eq:accFlux}
\end{gather}
Thus the inner-edge condition implies $F(r_{\rm ISCO})=0$. On the stable branches used here, $L'\geq0$ and $\Omega'<0$, giving nonnegative flux. In the far field $F\simeq3M\dot M/(8\pi r^3)$, with inner-edge-dependent subleading corrections. If all circular quantities are approximated by their Newtonian forms before integrating, the familiar factor $1-\sqrt{r_{\rm ISCO}/r}$ is also recovered. A finite-temperature maximum consequently occurs outside the ISCO rather than at the horizon.

For numerical work define $\omega=M\Omega$, $\ell=L/M$, and the dimensionless flux $\mathcal F=4\pi M^2F/\dot M$. Equation~\eqref{eq:accFlux} becomes
\begin{equation}
\mathcal F(x)=-\frac{d\omega/dx}{xH(x)}
\int_{x_{\rm ISCO}}^x\sqrt{H(s)}\frac{d\ell}{ds}\,ds.\label{eq:accFhat}
\end{equation}
No mass accretion rate is hidden in the plotted dimensionless profiles. With physical mass $M_{\rm phys}$ and mass supply $\dot M_{\rm phys}$, define the flux and temperature scales
\begin{gather}
F_*\equiv\frac{\dot M_{\rm phys}c^6}{4\pi G^2M_{\rm phys}^2},\label{eq:accTemp}\\
\shortintertext{This scale restores physical flux units through}
F_{\rm phys}=F_*\mathcal F,\label{eq:physicalFlux}\\
\shortintertext{The local Stefan-Boltzmann law gives the effective temperature:}
T_{\rm eff}=T_*\mathcal F^{1/4},\label{eq:effectiveTemperature}\\
\shortintertext{The temperature unit is fixed by}
T_*=(F_*/\sigma_{\rm SB})^{1/4}.\label{eq:temperatureScale}
\end{gather}
This follows from the local blackbody relation $F=\sigma_{\rm SB}T_{\rm eff}^4$. A color-correction factor is set to unity. Figure~16 compares the flux and temperature at a common physical radius, and then rescales the radius by each model's ISCO and the amplitude by its own peak. This separates the displacement of the emitting edge and changes in peak amplitude from changes in profile shape. Figure~17 displays stationary local-temperature maps and matched-radius NED/RN contrasts. No time evolution is implied by a sequence of stationary parameter choices.

\begin{figure*}[!tp]\centering\includegraphics[width=.94\linewidth,height=.78\textheight,keepaspectratio]{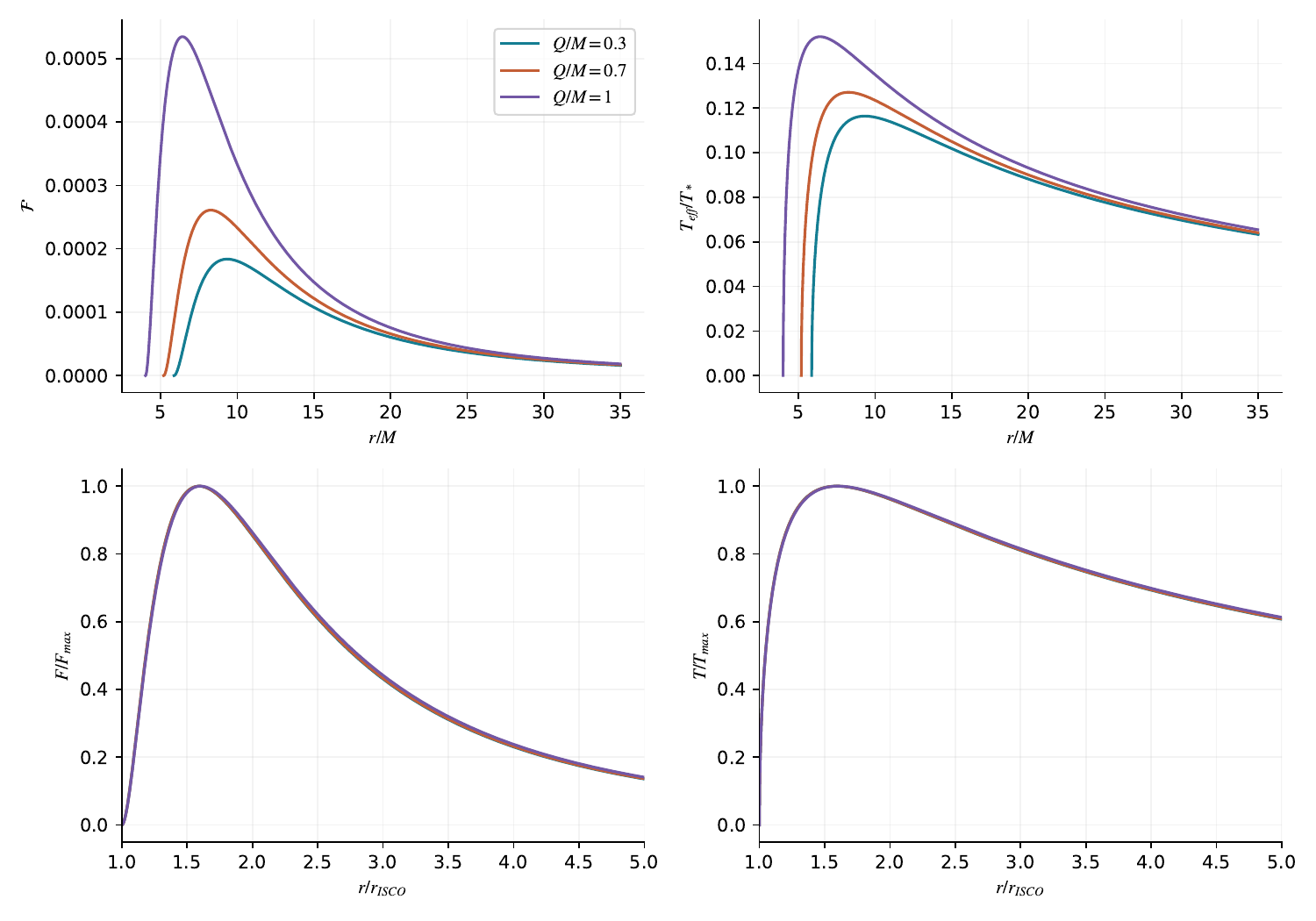}\caption{Zero-torque orbiting thermal-disk flux and temperature at $\widehat\mu=0.02$, measured in the gas frame. Upper panels use a common radius $r/M$ and show $\mathcal F=4\pi M^2F/\dot M$ and $T/T_*$. Lower panels use $r/r_{ISCO}$ and normalize each curve by its own flux or temperature peak. This comparison distinguishes the inner-edge shift and amplitude change from alterations of the radial shape. Flux and temperature vanish at each ISCO.}\end{figure*}

For illustration, $M_{\rm phys}=10^6M_\odot$ and $\dot M_{\rm phys}=10^{-4}M_\odot\,\mathrm{yr}^{-1}$ give $T_*\simeq7.770\times10^5\,$K. At fixed dimensionless metric parameters, $T_*\propto\dot M_{\rm phys}^{1/4}M_{\rm phys}^{-1/2}$. The example is a scaling of the stationary model, not a fit to a particular source.

\begin{figure*}[]\centering\includegraphics[width=.94\linewidth,height=.78\textheight,keepaspectratio]{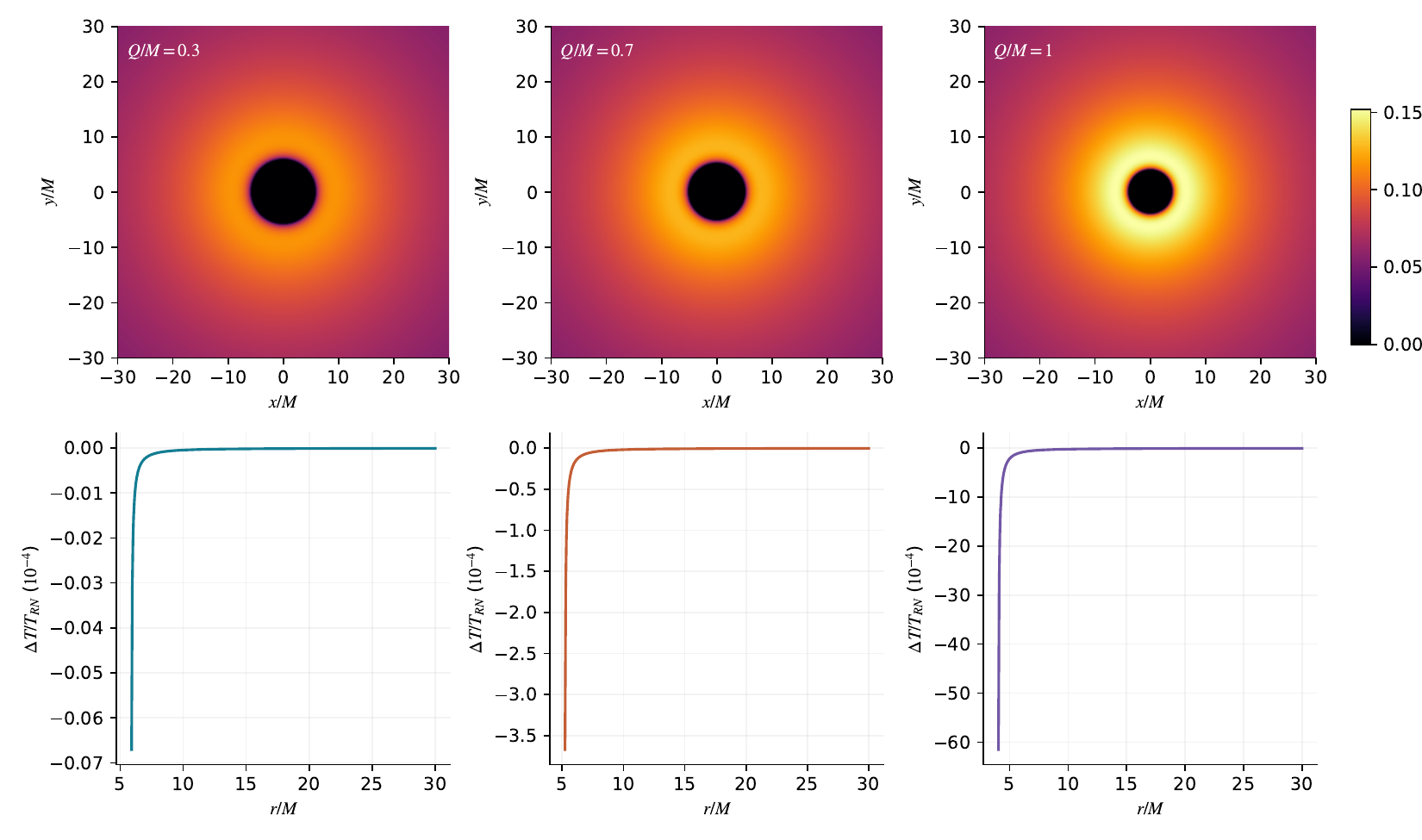}\caption{Stationary local-temperature maps of the circularly orbiting thermal disk for $Q/M=0.3,0.7,1.0$ at $\widehat\mu=0.02$. Stationarity concerns the temperature field, while the gas has nonzero orbital velocity. The upper maps show the intrinsic disk temperature $T/T_*$ in coordinate projections, with an emission-free central region inside the ISCO and a common color scale. They are not lensed observer images or a time sequence. Lower panels show the matched-radius fractional temperature change from RN, in units of $10^{-4}$, starting slightly outside the NED ISCO to avoid dividing by a zero edge flux. The temperature field is stationary.}\end{figure*}

\begin{figure*}[]\centering\includegraphics[width=.94\linewidth,height=.78\textheight,keepaspectratio]{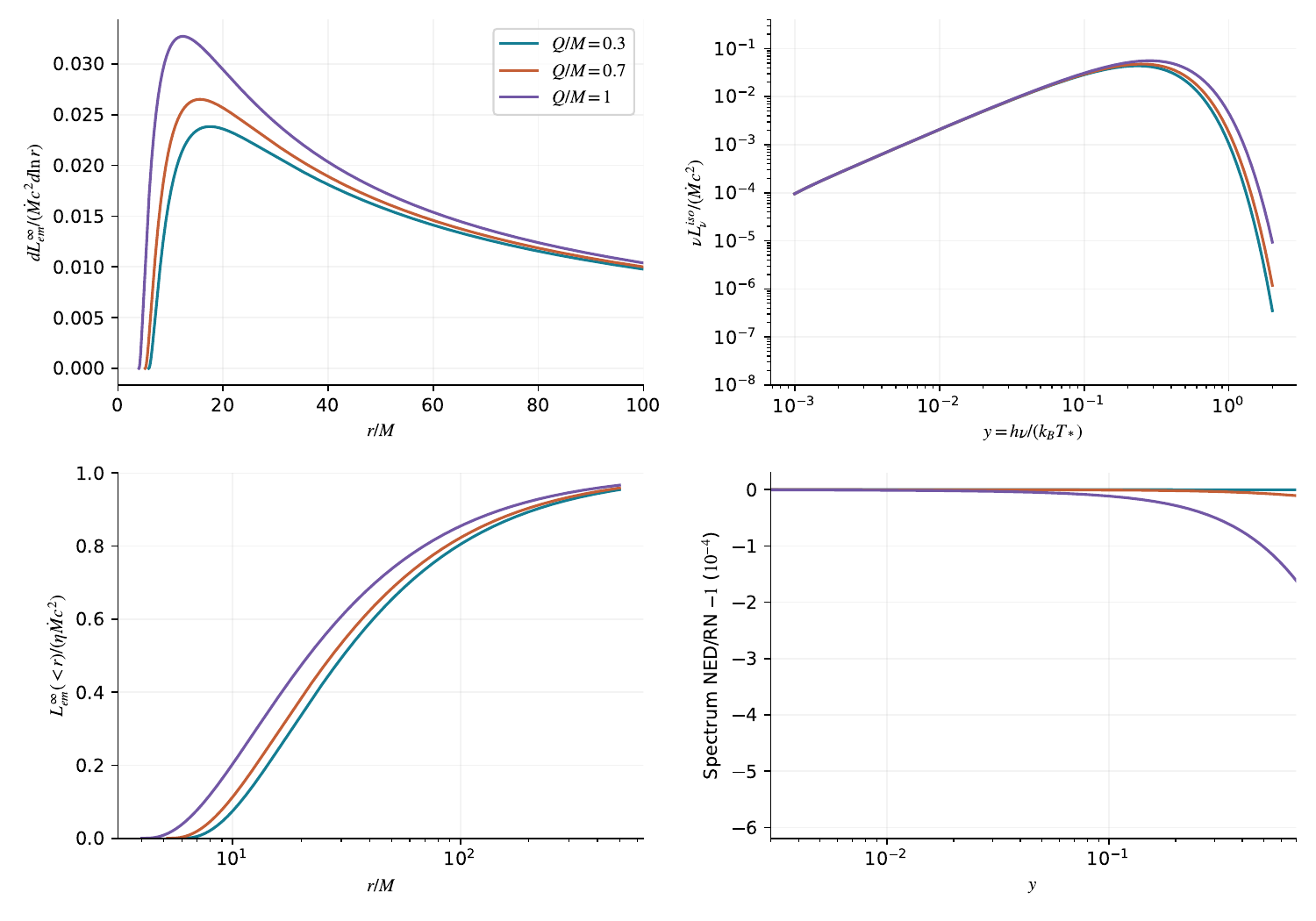}\caption{Radial luminosity and face-on continuum diagnostics for the orbiting thermal disk of Section~VI at $\widehat\mu=0.02$. Upper left: emitted Killing-energy luminosity per logarithmic radius, normalized by $\dot Mc^2$. Upper right: direct, unlensed isotropic-equivalent spectrum with the circular-emitter redshift, using $y=h\nu/(k_BT_*)$. Lower left: the fraction of emitted Killing energy enclosed within radius $r$. Lower right: the fractional continuum change relative to RN at the same charge, in units of $10^{-4}$. The two luminosity definitions are distinguished in the text; the continuum is not an angle-integrated escaping-energy budget.}\end{figure*}

The two-face emitted Killing-energy luminosity per logarithmic radius is
\begin{gather}
\frac{dL_{\rm em}^\infty}{d\ln r}=4\pi r^2E(r)F(r),\label{eq:accLum}\\
\shortintertext{The same emitted-energy distribution takes the dimensionless form}
\frac{1}{\dot M_{\rm phys}c^2}\frac{dL_{\rm em}^\infty}{d\ln x}=x^2E(x)\mathcal F(x).\label{eq:accLumHat}
\end{gather}
\begin{figure*}[!tp]\centering\includegraphics[width=.94\linewidth,height=.78\textheight,keepaspectratio]{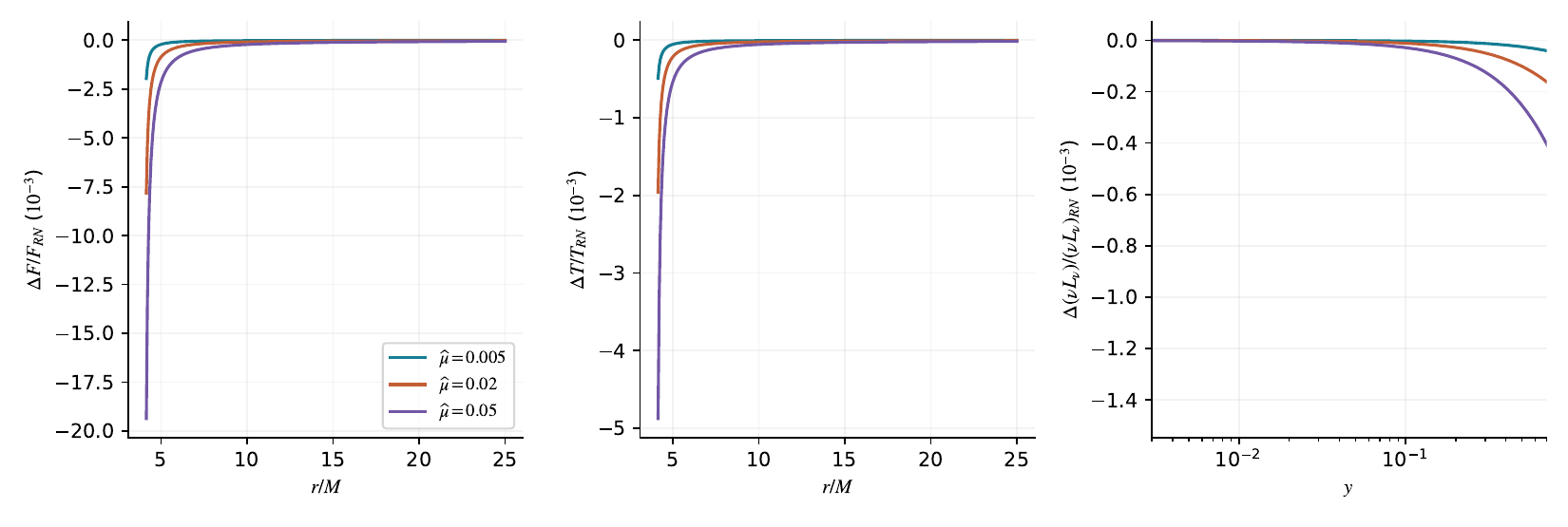}\caption{Coupling-only radiation diagnostics within the orbiting thermal-disk family at fixed $Q/M=1$. Flux, local temperature and the direct face-on continuum are compared with RN at the same radius or frequency, for $\widehat\mu=0.005,0.02,0.05$. All fractional changes are displayed in units of $10^{-3}$. Radii in the first two panels are restricted to the common emitting domain. These residuals resolve the small NED corrections that are hidden in the absolute disk profiles.}\end{figure*}

The superscript indicates Killing energy referred to infinity, not that every emitted photon escapes there. Integrating the energy conservation law with vanishing $\Omega\mathcal T$ at infinity gives
\begin{gather}
\frac{L_{\rm em}^\infty(<x)}{\dot M_{\rm phys}c^2}=\int_{x_{\rm ISCO}}^x sE(s)\mathcal F(s)\,ds,\label{eq:accBudget}\\
\shortintertext{With a zero-torque inner edge, the integral to infinity recovers the binding-energy budget~\cite{PageThorne}:}
L_{\rm em}^\infty(\infty)=\eta\dot M_{\rm phys}c^2.\label{eq:totalEmittedLuminosity}
\end{gather}

\begin{figure*}[!tp]\centering\includegraphics[width=.94\linewidth,height=.78\textheight,keepaspectratio]{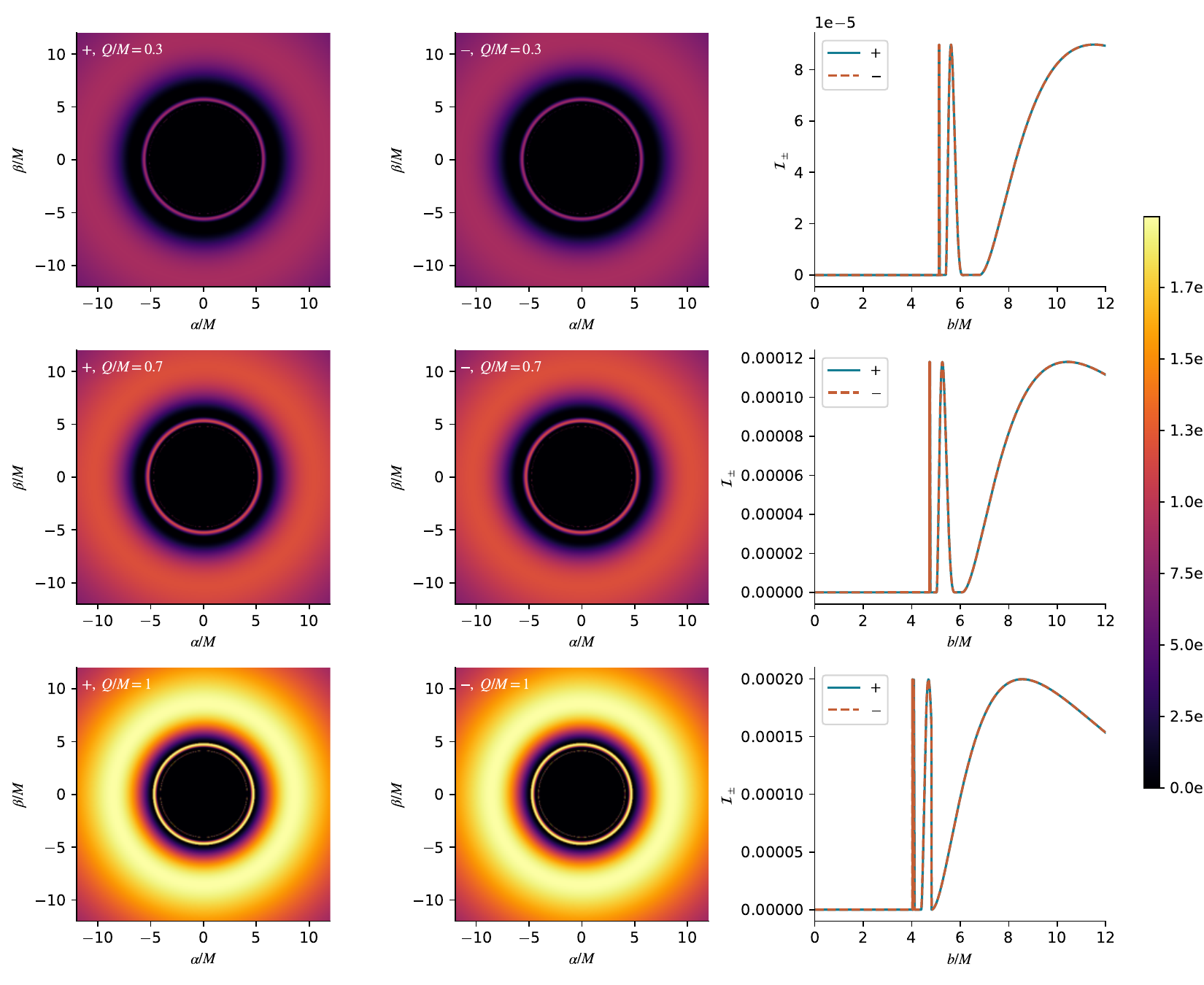}\caption{Thermal observer images of the circularly orbiting, optically thick disk using the two effective photon metrics, for $Q/M=0.3,0.7,1.0$ from top to bottom and $\widehat\mu=0.02$. This material and emission model differs from the static-source images of Section~III. Left column: plus mode. Middle column: minus mode. Right column: their radial bolometric intensity profiles. The disk is opaque and a backwards ray stops at its first visible intersection outside the ISCO, skipping crossings through the central empty hole. The local temperature is the same for both modes; the mapping and visible radius differ. The circular-emitter redshift is $g=\sqrt H$. All six maps share the displayed brightness scale in $\mathcal I_\pm=I_{\rm bol,\pm}^{obs}/[F_*/(2\pi)]$ and screen extent $[-12M,12M]$. This construction combines the thin-disk temperature calculation with polarization-dependent ray tracing.}\end{figure*}

The cumulative profile provides radii enclosing any chosen fraction of the emitted Killing energy. Table~\ref{tab:disk_diagnostics} gives the 50 and 90 percent radii, as well as the local flux peak, for the three charge examples. The flux integral, including its large-radius tail, reproduces $1-E_{\rm ISCO}$ within $3\times10^{-8}$ in these dimensionless examples.

\begin{table}[ht]
	\centering
	\small
	\caption{Stationary disk diagnostics at $\widehat\mu=0.02$. $x_F$ is the local flux/temperature peak radius; $x_{50}$ and $x_{90}$ enclose 50 and 90 percent of the emitted Killing-energy luminosity. Temperature examples use $M=10^6M_\odot$ and $\dot M=10^{-4}M_\odot\,\mathrm{yr}^{-1}$.}
	\label{tab:disk_diagnostics}
	\resizebox{\columnwidth}{!}{%
		\begin{tabular}{crrrrrr}
			\toprule
			$\widehat Q$ & $\eta$ & $x_F$ & $\mathcal F_{\max}$ & $T_{\max}$ (K) & $x_{50}$ & $x_{90}$ \\
			\midrule
			0.3 & 0.0584134 & 9.33265 & 0.000183726 & 90461 & 30.6017 & 213.622 \\
			0.7 & 0.0652557 & 8.26356 & 0.000261009 & 98760 & 27.2738 & 190.969 \\
			1.0 & 0.0814350 & 6.41132 & 0.000534594 & 118147 & 21.5570 & 152.305 \\
			\bottomrule
		\end{tabular}%
	}
\end{table}
\begin{figure*}[!tp]
	\centering
	\includegraphics[width=.94\textwidth, keepaspectratio]{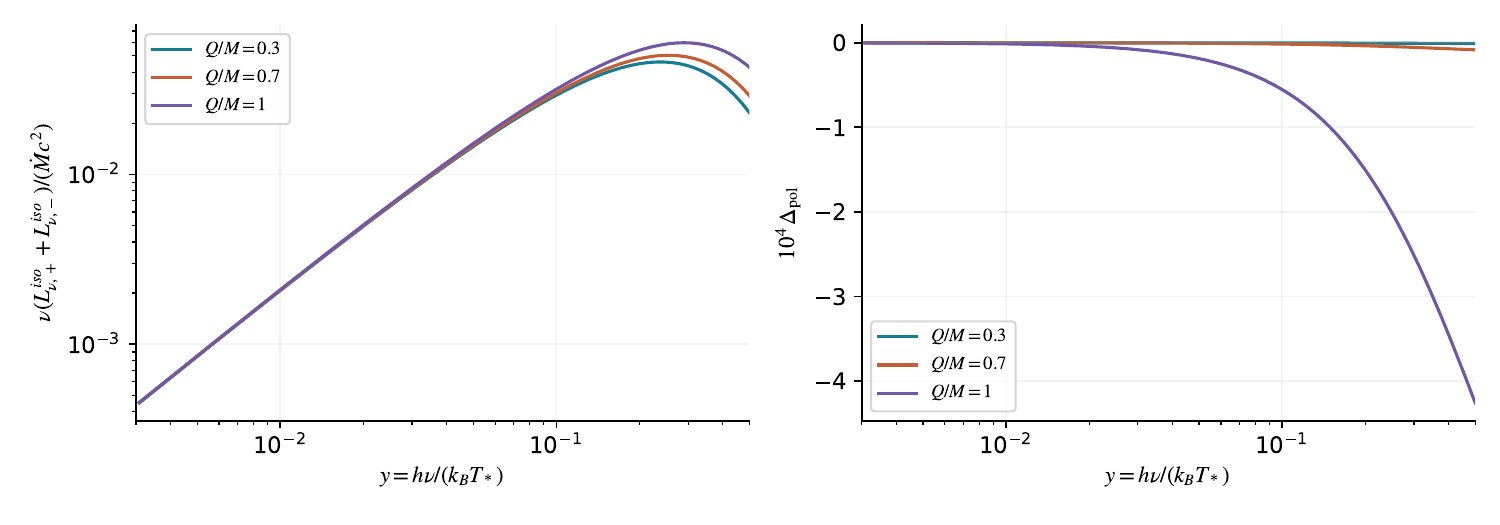}
	\caption{Compact thermal-spectrum comparison for $Q/M=0.3,0.7,1.0$ at $\widehat\mu=0.02$, using the orbiting opaque disk, circular-emitter redshift $g=\sqrt H$, and effective photon ray maps of Fig.~20. These spectra describe the Section~VI source rather than the static emitters of Sections~IIIIV. Left: the summed isotropic-equivalent spectrum $\nu(L_{\nu,+}^{iso}+L_{\nu,-}^{iso})/(\dot M c^2)$. Right: the signed fractional mode splitting $\Delta_{\rm pol}=2(L_{\nu,+}^{iso}-L_{\nu,-}^{iso})/(L_{\nu,+}^{iso}+L_{\nu,-}^{iso})$, displayed in units of $10^{-4}$. Positive values indicate a brighter plus mode, negative values a brighter minus mode, and zero equal mode spectra. This residual resolves the polarization difference without duplicating nearly coincident absolute curves. Each mode receives half the emitted unpolarized Planck intensity. The plotted interval $0.003\leq y\leq0.5$, with $y=h\nu/(k_BT_*)$, limits sensitivity to the finite outer edge. The screen integration extends to $\xi=10^5$ and the local disk profiles to $x=10^6$. The spectra include ray mapping and differ from the direct continuum approximation in Fig.~18. No additional NED amplitude-transport correction is assumed.}
	\label{fig:disk8_compact}
\end{figure*}

The peak and enclosing radii decrease with increasing charge in this comparison, while the peak temperature and efficiency increase. The positive NED correction at the small couplings used here is far smaller than the charge dependence. Figures~1819 include RN-subtracted flux, temperature and spectral ratios so those two effects can be assessed separately.

A rotating emitter has four-velocity $u^\alpha=u^t(1,0,0,\Omega)$. If $\lambda_z=p_\phi/(-p_t)$ is the photon's axial angular-momentum ratio, its frequency shift to an observer at infinity follows the standard circular-emitter transfer relation~\cite{Cunningham,Li}:
\begin{equation}
g=\frac{\nu_{\rm obs}}{\nu_{\rm em}}=\frac{1}{u^t(1-\Omega\lambda_z)}.\label{eq:accRedshift}
\end{equation}
For the polar observer, axial symmetry gives $\lambda_z=0$, hence $g_{\rm circ}=\sqrt H$. Circular motion still contributes transverse Doppler redshift in this face-on geometry. In the local static frame, $v_{\rm orb}^2=r^2\Omega^2/f=rf'/(2f)$, so $g_{\rm circ}=\sqrt f\sqrt{1-v_{\rm orb}^2}$. The earlier static emitters instead have $v_{\rm orb}=0$ and $g_{\rm stat}=\sqrt f$. These shifts follow from the different material four-velocities; changing between them changes the source configuration.

For local isotropic blackbody emission in the orbiting gas frame, $B_\nu(T)=2h\nu^3/[c^2(\exp(h\nu/k_BT)-1)]$. We apply the same geometric-optics intensity invariant as earlier, now with the thermal source and its circular-emitter shift: $I_\nu^{\rm obs}=g_{\rm circ}^3B_{\nu/g_{\rm circ}}(T)=B_\nu(g_{\rm circ}T)$. Sharing this transfer invariant does not imply a shared material or emission model across sections. A simple face-on continuum approximation neglecting lensing sets the screen impact parameter equal to the emitting cylindrical radius. The redshifted-blackbody construction~\cite{Cunningham,Li} gives the face-on observer flux
\begin{gather}
F_{\nu,\rm obs}=\frac{2\pi}{D^2}\int_{r_{\rm ISCO}}^\infty B_\nu(gT)r\,dr,\label{eq:accSpec}\\
\shortintertext{We express this face-on flux as an isotropic-equivalent spectral luminosity through}
L_\nu^{\rm iso}=4\pi D^2F_{\nu,\rm obs}.\label{eq:isotropicLuminosity}
\end{gather}
The factor $4\pi D^2$ defines the isotropic-equivalent luminosity; a disk is not an isotropic radiator. This quantity is distinct from the emitted Killing-energy luminosity in Eq.~\eqref{eq:accLum}. With $y=h\nu/(k_BT_*)$, use the blackbody StefanBoltzmann constant to obtain the dimensionless spectrum
\begin{equation}
\frac{\nu L_\nu^{\rm iso}}{\dot M_{\rm phys}c^2}
=\frac{30}{\pi^4}y^4\int_{x_{\rm ISCO}}^\infty
\frac{x\,dx}{\exp\left[y/(\sqrt{H(x)}\mathcal F(x)^{1/4})\right]-1}.\label{eq:accSpecHat}
\end{equation}
The integrand vanishes at the zero-flux inner edge. Equations~\eqref{eq:accSpec}\eqref{eq:accSpecHat} define the frequency axis and continuum normalization explicitly. Figure~18 juxtaposes this spectrum with the differential and cumulative emitted-energy distributions; the differing definitions are kept explicit. Figure~19 isolates the effect of varying $\widehat\mu$ at fixed $Q/M=1$.

We additionally construct the two photon transfer maps from the effective photon metrics of the adopted model to calculate polarization-dependent thermal images and spectra. At each background-parameter choice, the orbiting material disk and its $F,T,E,L$ are identical for the two polarization modes. This source matching is internal to the thermal calculation. The disk is optically thick in this thermal calculation: a backwards ray stops at its first intersection with $r\geq r_{\rm ISCO}$, denoted $r_{\rm vis}^\pm(b)$. Intersections inside the central disk-free hole are skipped. This differs from the optically thin sum of multiple emitting intersections used for the earlier prescribed source models. Each emitting face carries half its unpolarized blackbody intensity in each mode.

For $\xi=b/M$, define $\vartheta_\pm(\xi)=\sqrt{H(r_{\rm vis}^\pm)}\mathcal F(r_{\rm vis}^\pm/M)^{1/4}$; set it to zero if no disk intersection is visible. The thermal images and mode spectra are
\begin{gather}
\mathcal I_\pm(\xi)\equiv\frac{I_{\rm bol,\pm}^{\rm obs}}{F_*/(2\pi)}=\vartheta_\pm(\xi)^4,\label{eq:accMode}\\
\shortintertext{Integrating the redshifted Planck intensity over the screen gives the spectrum of each mode:}
\frac{\nu L_{\nu,\pm}^{\rm iso}}{\dot M_{\rm phys}c^2}=\frac{15}{\pi^4}y^4\int_0^\infty\frac{\xi\,d\xi}{\exp[y/\vartheta_\pm(\xi)]-1}.\label{eq:modeSpectrum}
\end{gather}
The two mode spectra sum to the unresolved thermal spectrum. Their coefficient is half the unpolarized expression because the emission is split equally. Screen integration incorporates the different ray mappings of the two metrics rather than setting $b=r$. The photon factors affect the visible-emission radius and lensing; the gas clock remains the background circular-orbit $u^t$. These spectra characterize the orbiting thermal source, whereas Sections~IIIIV characterize prescribed static emitters with different emission and opacity assumptions.

Figure~20 shows the polarization-sensitive thermal images and radial intensities. Figure~21 presents their summed thermal spectra and the signed difference normalized by the mean mode spectrum, making the small polarization splitting visible in a compact comparison. Within this orbiting thermal-disk family, the fixed-charge RN residuals in Figure~19 isolate the intrinsic nonlinear-coupling correction, while Figure~21 compares propagation through the two effective photon metrics using the same material disk. As in the earlier optical calculations, these adopt the ordinary $I_\nu/\nu^3$ transport prescription; the material four-velocity and emissivity remain those of the present thermal source. Brightness differences between these thermal images and the static-source images in Section~III cannot be assigned solely to photon propagation, because the source configurations differ. NED mode-amplitude transport, atmosphere color correction, absorption, returning irradiation and self-consistent disk MHD are not included. The ray tracing determines which surface is visible but does not feed captured or returning radiation back into the local dissipation model. The figures therefore distinguish intrinsic disk heating, emitted Killing energy and an approximate observer spectrum, instead of treating all three as one luminosity.


\section{Conclusion}

We have studied birefringent photon propagation, interferometric visibility, neutral bound motion, and stationary disk radiation within a common magnetically charged ENED background with the EH-type Lagrangian and free coupling adopted from de Paula et al.~\cite{dePaula}. The radiative calculations use distinct material-source configurations: prescribed static, optically thin emitters in Sections~IIIIV and a circularly orbiting, optically thick thermal disk in Section~VI. The two effective photon geometries produce distinct unstable light rings and capture thresholds throughout the sampled admissible exterior domain. Increasing magnetic charge moves the critical optical scales inward in the displayed comparisons, whereas the sampled nonlinear coupling shifts them outward and separates the two polarization sectors. For each prescribed source and fixed background parameters, paired polarization images and transfer profiles resolve the small optical separation while using the same material source in both modes. The observed brightness peaks also depend on the emissivity and image order, so they cannot be identified directly with the capture boundary.

For the specified static, optically thin source, the interferometric extension separates direct, lensed, and higher-order emission in each polarization image and transforms their brightness into signed visibility signals. Their unresolved sum admits a narrow-ring modulation scale set by the separation of the two critical curves. The small separation in the examples makes this scale substantially longer than in a black hole with two widely separated photon spheres. This correspondence concerns the Fourier signal: photons do not move between barriers belonging to different polarization sectors. Finite subring widths and source structure remain present in the numerical image, so the analytic beat envelope does not by itself establish a measurable signature.

The convergence test supports the eight-order photon sum on $3000\leq u\leq25000$ through comparisons of the signed visibility and slow-modulation minima with sums through order 16 and independent quadrature refinement. The ideal thin-ring extrapolation has a separate limitation: using unexpanded rational optical factors changes the beat spacing by approximately $0.9\%$ and $7.4\%$ relative to its leading-order value for the two displayed charges. These are prescription-sensitivity estimates, not higher-order theoretical error bounds. At $Q/M=1$, horizon-sensitive emissivities additionally inherit the nonanalytic $\sqrt{\widehat\mu}$ horizon response, including the model-3 source normalization; this effect must be distinguished from the regular exterior polarization splitting.

The massive-particle calculation supplies a complementary probe of the background geometry. Closed trajectories with a common rational orbital topology move inward and acquire shorter radial periods as charge increases in the examples studied. Their gravitational-wave snapshots reflect the resulting timing and polarization changes, while Maxwell-subtracted comparisons expose the smaller additional nonlinear contribution. The admissible energyangular-momentum domain distinguishes stable bound motion from the separatrix and near-unbound limit. These trajectories and snapshots are conservative reference calculations; radiation reaction and consistent perturbation dynamics are needed to turn them into evolving inspiral waveforms.

The thin-disk analysis consistently relates circular motion, angular-momentum transport, radiative flux, temperature, and emitted binding-energy efficiency. In the displayed charge sequence, the emitting region becomes more compact, hotter, and more efficient. Comparisons at the same charge show that the nonlinear-electrodynamic changes are much smaller than the charge-driven shifts in the explored weak-coupling range. For each thermal-disk model, applying both effective ray maps to the same orbiting material source gives polarization-resolved images and spectra that distinguish intrinsic heating from mode-dependent transfer. This source matching applies within Section~VI; it does not make the thermal disk the material configuration used for the earlier static-source images and visibilities. Emitted efficiency, isotropic-equivalent spectra, and capture-corrected escaping luminosity remain different quantities.

These calculations provide complementary diagnostics of the optical and background geometry under their separately stated material assumptions. The static-source brightness and visibility results and the orbiting-disk thermal spectra are not alternative predictions for one fixed accretion configuration. Such a prediction would require a common velocity field, emissivity, opacity, and boundary prescription across the radiation calculations. Quantitative interpretation also remains conditional on the truncated electromagnetic theory, the adopted geometric-optics intensity prescription, the stationary zero-torque blackbody disk, and the quadrupole snapshot approximation. More realistic amplitude transport, disk atmospheres, returning radiation, and dissipative orbital evolution would be needed before drawing observational constraints from the predicted differences.

\Needspace{6\baselineskip}
\section*{\texorpdfstring{Acknowledgments}{Acknowledgments}}
S.H.Dong thanks the partial support of the project 20260193-SIP-IPN, Mexico. OL thanks Roberto Della Ceca and the Brera National Institute of Astrophysics for support and acknowledges the Istituto Nazionale di
Fisica Nucleare (INFN) through the Iniziativa Specifica
QGSKY. 

\vspace{0.25\baselineskip}

\end{document}